\documentclass[a4paper,fleqn]{cas-dc}

\usepackage[authoryear]{natbib}
\usepackage{graphicx}
\usepackage{amsmath,amssymb,amsfonts}
\usepackage{booktabs}
\usepackage{xcolor}
\usepackage{hyperref}
\usepackage{xurl}
\usepackage{placeins}
\usepackage{stfloats}

\begin{document}

\shorttitle{AtlasPatch: Efficient \& Scalable Slide Preprocessing}
\shortauthors{A. Alagha et~al.}



\title [mode = title]{{AtlasPatch}: Scalable Foundation Model-based Tissue Detection and Patch Extraction for Computational Pathology}

\tnotemark[1]

\tnotetext[1]{\textbf{The AtlasPatch package is available on \url{https://github.com/AtlasAnalyticsLab/AtlasPatch}}}

\author[1,2]{Ahmed Alagha}\cormark[1]
\ead{ahmed.alagha@mail.concordia.ca}
\credit{Conceptualization, Data curation, Methodology, Project administration, Validation, Writing - Original draft preparation}

\author[1]{Christopher Leclerc}
\ead{christopher.leclerc@mail.concordia.ca}
\credit{Data curation, Methodology, Validation}

\author[1,5]{Yousef Kotp}\fnref{equal}
\ead{yousef.kotp@mila.quebec}
\credit{Methodology, Validation, Software, Writing - review and editing}

\author[1]{Omar Metwally}\fnref{equal}
\ead{omar.abdelwahed@mail.concordia.ca}
\credit{Methodology, Validation, Software, Writing - review and editing}

\author[1]{Calvin Moras}
\ead{calvin.moras@mail.concordia.ca}
\credit{Data curation}

\author[1]{Peter Rentopoulos}
\ead{peter.rentopoulos@mail.concordia.ca}
\credit{Data curation}

\author[1,3]{Ghodsiyeh Rostami}
\ead{rose.rostami@mail.concordia.ca}
\credit{Methodology, Writing - Original draft preparation}

\author[4]{Bich Ngoc Nguyen}
\ead{bich.ngoc.nguyen.med@ssss.gouv.qc.ca}
\credit{Data curation, Resources}

\author[4]{Jumanah Baig}
\ead{jumanah.baig@mail.concordia.ca}
\credit{Data curation, Resources}

\author[4]{Abdelhakim Khellaf}
\ead{abdelhakim.khellaf@umontreal.ca}
\credit{Data curation, Resources}

\author[4,8]{Vincent Quoc-Huy Trinh}
\ead{quoc-huy.trinh@umontreal.ca}
\credit{Data curation, Resources, Validation, Writing -- review and editing}

\author[6]{Rabeb Mizouni}
\ead{rabeb.mizouni@ku.ac.ae}
\credit{Supervision, Writing -- review and editing}

\author[6]{Hadi Otrok}
\ead{hadi.otrok@ku.ac.ae}
\credit{Supervision, Writing -- review and editing}

\author[6,2]{Jamal Bentahar}
\ead{jamal.bentahar@ku.ac.ae}
\credit{Supervision, Funding acquisition, Writing -- review and editing}

\author[1,5,7]{Mahdi S. Hosseini}
\ead{mahdi.hosseini@concordia.ca}
\credit{Conceptualization, Data curation, Methodology, Project administration, Funding acquisition, Supervision, Validation, Writing -- review and editing}

\affiliation[1]{organization={Department of Computer Science and Software Engineering, Concordia University},
                addressline={2155 Guy St},
                postcode={H3G 1M8},
                city={Montreal, QC},
                country={Canada}}

\affiliation[2]{organization={Concordia Institute for Information Systems Engineering, Concordia University},
                addressline={1515 Rue Ste-Catherine O},
                postcode={H3G 1S6},
                city={Montreal, QC},
                country={Canada}}

\affiliation[3]{organization={Department of Building, Civil, and Env. Engineering, Concordia University},
                addressline={1515 Rue Ste-Catherine O},
                postcode={H3G 1S6},
                city={Montreal, QC},
                country={Canada}}

\affiliation[4]{organization={University of Montreal Hospital Center (CHUM)},
                addressline={1000 R. Saint-Denis},
                postcode={H2X 0C1},
                city={Montreal, QC},
                country={Canada}}

\affiliation[5]{organization={Mila--Quebec AI Institute},
                addressline={6666 Rue Saint-Urbain},
                postcode={H2S 3H1},
                city={Montreal, QC},
                country={Canada}}

\affiliation[6]{organization={Department of Computer Science, Khalifa University},
                addressline={Shakhbout Bin Sultan St},
                city={Abu Dhabi},
                country={United Arab Emirates}}

\affiliation[7]{organization={Department of Pathology, McGill University},
                addressline={3775 University Street},
                postcode={H3A 2B4},
                city={Montreal, QC},
                country={Canada}}

\affiliation[8]{organization={Institute for Research in Immunology and Cancer, University of Montreal},
                addressline={2950, Chemin de Polytechnique},
                postcode={H3T 1J4},
                city={Montreal, QC},
                country={Canada}}

\cortext[cor1]{Corresponding author}
\fntext[equal]{These authors contributed equally to this work.}

\begin{abstract}
Whole-slide image (WSI) preprocessing, including tissue detection and patch extraction, is a critical step in computational pathology, yet remains a major bottleneck for large-scale workflows. Existing preprocessing methods often rely either on threshold-based heuristics that are sensitive to staining variations, tissue fragmentation, and artifacts, or on patch-wise deep learning pipelines with substantially higher computational cost. We present AtlasPatch, a scalable high-throughput WSI preprocessing method built around a foundation-model-based tissue detector that operates at thumbnail resolution: a single thumbnail-level forward pass yields a tissue mask that directly guides patch coordinate generation at the target desired magnification, avoiding repeated patch-level inference. The proposed detector's robustness and efficiency is driven by two coupled contributions: (i) a parameter-efficient adaptation of the SAM2 foundation model that updates only its layer-normalization parameters ($\sim$0.076\% of model weights), and (ii) a curated and semi-manually annotated multi-cohort dataset of $\sim$30,000 WSI thumbnail–mask pairs deliberately spanning multiple organs, scanners, tissue appearances, and artifacts. The detector is coupled with pyramid-aware contour mapping from thumbnail to full-resolution slide coordinates, enabling direct patch coordinate generation at the target magnification and parallelized high-throughput patch extraction. AtlasPatch's tissue detection achieves a precision of 0.986 and remains robust across variations in brightness, fragmentation, boundary definition, tissue heterogeneity, and common artifacts. Compared with widely used deep-learning preprocessing methods, AtlasPatch is up to 16$\times$ faster while preserving downstream multiple-instance learning performance across six slide-level classification tasks. These results position AtlasPatch as a frontier of efficient preprocessing in large-scale computational pathology and pathology foundation models.

\end{abstract}



\maketitle

\section{Introduction}\label{sec:intro}

Modern pathology is increasingly transitioning from conventional microscopy to digital workflows enabled by whole slide images (WSIs), which are high-resolution digital scans of conventional glass tissue slides \citep{Hanna2022Integrating}. WSIs support scalable archiving, rapid retrieval, and seamless multi-resolution viewing from slide-level context to cellular detail \citep{Zia2025Update}, and these capabilities helped unlock modern deep learning (DL) for tasks such as tumor detection, grading, prognosis, and biomarker discovery \citep{hosseini2024computational}. Because gigapixel WSIs exceed the practical limits of most neural network architectures, with a large fraction of each slide being non-tissue background, analysis typically proceeds on small tissue patches, with slide-level diagnoses inferred via multiple-instance learning (MIL) or other aggregation methods \citep{Lu2021CLAM, zhang20252dmamba, gadermayr2024multiple, hosseini2026efficient, li2025diagnostic}. Naive tiling of such large-scale WSIs produces millions of candidate patches per cohort, much of it being background, inflating I/O, memory and wall-clock time. The scaling demands are magnified by the emergence of pathology foundation-models which require billions of diverse patches spanning organs, stains, scanners, and magnifications to achieve robustness \citep{dippel2024rudolfv}. Efficient tissue detection coupled with high-throughput patch extraction is therefore not merely an implementation detail but a hard constraint on the pace, cost, and reach of computational pathology.

Current slide-preprocessing pipelines detect tissue then tessellate WSIs into fixed-size patches for downstream tasks \citep{Lu2021CLAM, CLAMGit, Zhang2025Trident, TRIDENTGit, pocock2022tiatoolbox, DPLabPaper, DPLabDocs}. Tissue detection is typically implemented either with simple heuristics (e.g., color thresholding and morphological filtering) or with deep-learning segmenters like U-Net \citep{Ronneberger2015UNet} and U-Net++ \citep{zhou2018unet++}. Thresholding-based tissue detectors, such as HistoQC \citep{Janowczyk2019HistoQC, HistoQCGit}, dplabtools \citep{DPLabPaper, DPLabDocs}, EntropyMasker \citep{EntropyMasker, EntropyMaskerGit}, TIAToolbox \citep{pocock2022tiatoolbox} and the WSI processor in CLAM \citep{Lu2021CLAM, CLAMGit} often require manual tuning across different cohorts by setting appropriate thresholds, and commonly fail under stain/tissue variations or artifacts despite being computationally fast. Deep learning-based pipelines, such as PathML \citep{Berman2021PathML, PathMLDocs} and TRIDENT \citep{Zhang2025Trident, TRIDENTDocs, TRIDENTGit} improve robustness via patch-level inference to identify tissue and background patch images, but scale poorly (hundreds--thousands of forward passes per WSI). Patch-wise detection can also miss global slide context (tissue geometry, background structure), leading to inconsistent boundaries after stitching. Moreover, many of the existing pipelines offer limited end-to-end parallelization, with several modules (such as WSI fetching, patch coordinate extraction, patch image writing) executed serially or with only coarse-grained parallelism. In practice, these factors make ``standard'' slide preprocessing acceptable in small studies, but increasingly expensive at foundation-model scale.

\begin{figure*}[pos=!htbp,width=\textwidth]
    \centering
    \includegraphics[width=\textwidth]{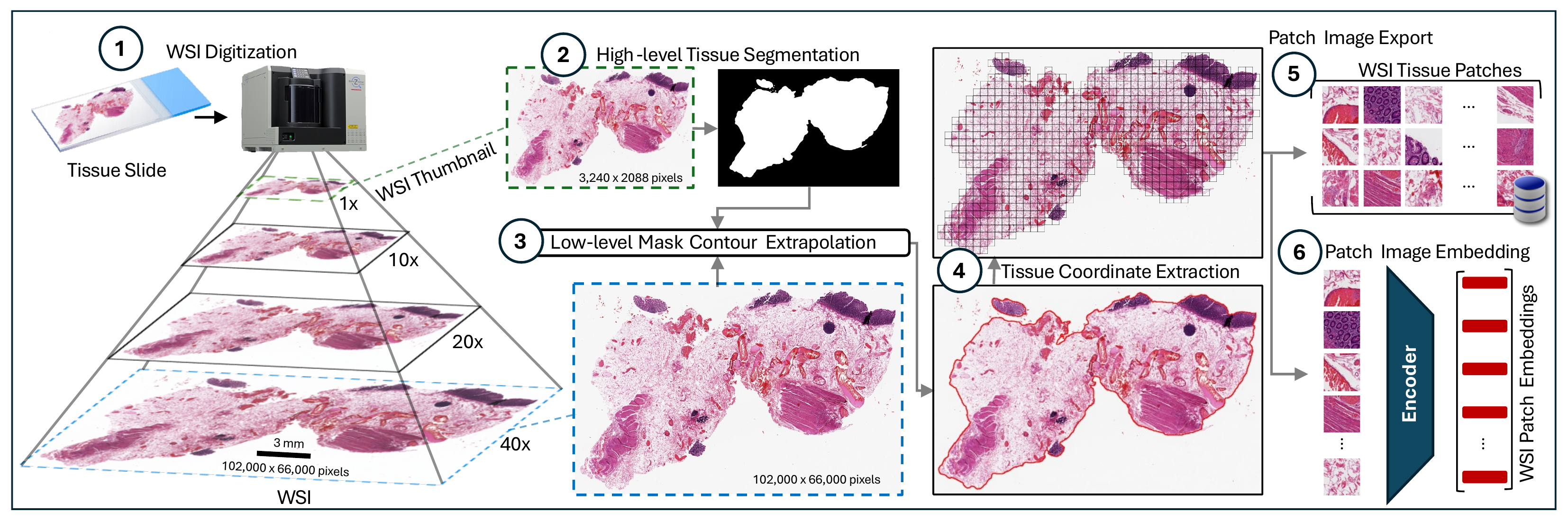}
    \caption{Overview of the AtlasPatch slide-preprocessing pipeline. Following the acquisition of diagnostic WSIs as multi-resolution pyramids, AtlasPatch utilizes the thumbnail for tissue detection using the finetuned SAM2 model, then extrapolates the resulting contour into the desired high resolution. This is followed by computing patch coordinates for downstream patch image export or feature embeddings.}
    \label{fig:overview}
\end{figure*}

Here we introduce AtlasPatch (Fig.~\ref{fig:overview}), a scalable high-throughput slide-preprocessing method that addresses key bottlenecks in tissue detection and patch extraction while preserving downstream performance. Rather than relying on heuristic thresholding or computationally intensive patch-wise inference, AtlasPatch performs tissue detection at thumbnail resolution using a parameter-efficiently adapted Segment Anything 2 (SAM2) \citep{ravi2024sam} foundation model, then transfers the resulting contours to full-resolution for patch coordinate extraction at the target magnification. To enable robust thumbnail-level tissue detection, we curated a large and deliberately heterogeneous multi-cohort set of WSI thumbnails and semi-manually annotated tissue masks spanning multiple organs, scanners, tissue appearances, and common artifacts. Building on this dataset, we develop a selective fine-tuning strategy for SAM2 that updates only the normalization parameters, substantially reducing training cost while maintaining strong segmentation performance. AtlasPatch is further designed as a parallelized end-to-end preprocessing framework, combining GPU-batched thumbnail inference with contour-guided patch coordinate generation, multi-core CPU parallelization for coordinate extraction and patch I/O, and optional GPU-accelerated embedding for downstream tasks. AtlasPatch produces clean thumbnail-level tissue masks that retain informative tissue and suppress common artifacts (e.g., pen marks). Across several public and in-house multi-organ classification tasks (kidney, lung, breast, colorectal), the full proposed framework achieves downstream performance comparable to, and sometimes exceeding, widely used preprocessing methods, while reducing runtime by up to 16-fold. The contributions of this work are summarized below:

\begin{itemize}
    \item The curation and semi-manual annotation of a large heterogeneous multi-cohort dataset of WSI thumbnails, providing robust supervision for thumbnail-level tissue detection across diverse organs, scanners, tissue appearances, and artifacts. 
    \item A parameter-efficient adaptation strategy for SAM2 that updates only a small subset of model parameters, substantially reducing training cost while maintaining strong segmentation performance.
    \item The design of AtlasPatch as a parallelized end-to-end preprocessing framework in which thumbnail-level tissue contours are transferred to full-resolution slide space to guide efficient patch coordinate generation and extraction, enabling high-throughput WSI processing without repeated patch-wise tissue inference.
    \item The extensive evaluation across stratified subsets of the dataset and diverse slide-level classification tasks, demonstrating strong adaptability across heterogeneous WSI conditions and competitive performance against several existing preprocessing methods.
    \item The release of AtlasPatch as an accompanying open-source preprocessing tool for the research and clinical pathology communities, with a modular design supporting tissue detection, patch coordinate extraction, embedding generation, and optional patch export.
\end{itemize}

Together, these design choices make AtlasPatch a scalable and efficient backbone for foundation-model-scale pathology workflows, where preprocessing can otherwise dominate runtime and cost. The remainder of this paper is organized as follows. Section \ref{sec:related} reviews related works in tissue detection and patch extraction, Section \ref{sec:method} presents the proposed method, Section \ref{sec:experiments} describes the experimental setup and analyzes the extensive results, Section \ref{sec:discussion} presents several views and discussions on the results, Section \ref{sec:conclusion} concludes with final remarks, and Section \ref{sec:availability} discusses data and code availabilities, with the Appendix section showing additional supporting figures.

\section{Related Works}\label{sec:related}
Related work in WSI preprocessing mainly falls into two categories: standalone tissue detection methods and end-to-end slide processing frameworks that combine tissue detection and patch extraction. This section covers related works in both categories.

\subsection{Tissue detection}\label{sec:related_tissue}

Accurate delineation of tissue from background is a prerequisite for virtually every computational pathology workflow, yet the task presents unique challenges due to the heterogeneity of whole-slide images across institutions, scanners, and staining protocols. Early and still widely adopted approaches rely on classical image processing heuristics. Otsu thresholding \citep{otsu1979threshold} on grayscale or HSV-converted thumbnails is the most common strategy, employed by pipelines such as the CLAM preprocessing module \citep{Lu2021CLAM, CLAMGit} and TIAToolbox's \texttt{TissueMasker} \citep{pocock2022tiatoolbox, TIADocs}. HistoQC \citep{Janowczyk2019HistoQC, HistoQCGit} extends thresholding with morphological filtering and color-variance checks to suppress adipose-like regions and small artifacts. Similarly, dplabtools \citep{DPLabPaper, DPLabDocs} generates binary masks via thresholding in HSV/Lab space with optional hole filling and small-object pruning, and EntropyMasker \citep{EntropyMasker, EntropyMaskerGit} uses a Shannon-entropy criterion on local image patches to separate tissue from background. While computationally efficient, these heuristic methods often require manual tuning of thresholds for each cohort and commonly fail under staining variation, low tissue--background contrast, or the presence of artifacts such as pen markings and scanner streaks \citep{Bandi2017Comparison, Bandi2019, ciga2021patchlimit}.

Deep learning approaches have been explored to improve robustness. Convolutional architectures such as U-Net \citep{Ronneberger2015UNet} and U-Net++ \citep{zhou2018unet++} have been applied to tissue segmentation at the patch level, and multi-resolution networks have been proposed to handle the scale diversity inherent in WSIs \citep{gu2018multi}. More recently, the Segment Anything Model (SAM) \citep{Kirillov2023SAM} and its successor SAM2 \citep{ravi2024sam} have demonstrated strong zero-shot segmentation on natural images, motivating adaptation to histopathology \citep{Zhang2023SAMPath, liu2024wsi}. The TRIDENT toolkit \citep{Zhang2025Trident, TRIDENTGit} provides two deep learning tissue detectors: one based on GrandQC \citep{weng2024grandqc}, which classifies thumbnail-level patches, and one based on Hest \citep{jaume2024hest}, which operates on higher-magnification (10$\times$) patches. While these patch-level methods improve accuracy, they incur substantially higher computational cost, requiring hundreds to thousands of forward passes per slide, and can lose global context when stitching patch-level predictions back into a slide-level mask. Interactive or semi-automatic annotation tools such as ilastik \citep{Sommer2011} and QuPath \citep{Bankhead2017QuPath} offer flexible tissue delineation but require per-slide human input, precluding their use in high-throughput pipelines. Parameter-efficient fine-tuning strategies, including Layer Normalization tuning \citep{zhaotuning, rostami2025segmentcrackdeepsemantic}, have recently been shown to effectively adapt large foundation models to domain-specific segmentation tasks with minimal parameter updates, offering a promising path for efficient tissue detection in histopathology.

\subsection{End-to-end WSI preprocessing pipelines}\label{sec:related_tools}

Beyond tissue detection alone, several frameworks aim to provide end-to-end preprocessing for computational pathology, encompassing slide loading, tissue detection, patch extraction, and feature encoding. The slide processor in CLAM \citep{Lu2021CLAM, CLAMGit} is among the most widely used: it couples Otsu-based tissue detection with contour-guided patch extraction and an attention-based MIL aggregator, but its heuristic tissue masking limits robustness on heterogeneous cohorts. TIAToolbox \citep{pocock2022tiatoolbox, TIADocs} offers a modular library with built-in tissue masking and patch extraction utilities, integrated into PyTorch-based analysis pipelines. However, its tissue detection similarly relies on Otsu and morphological heuristics. PathML \citep{Berman2021PathML, PathMLDocs} provides a broader end-to-end framework including quality control, stain normalization, and deep learning--based tissue detection, though its patch-level inference model adds significant computational overhead. TRIDENT \citep{Zhang2025Trident, TRIDENTGit, TRIDENTDocs} is a recent toolkit that benchmarks multiple tissue detectors and integrates them with patch extraction and embedding pipelines, but its deep learning detectors operate at the patch level, leading to runtime that scales linearly with slide size. Other tools include Slideflow \citep{dolezal2024slideflow, SlideflowGit}, which provides a unified deep learning pipeline for digital pathology from preprocessing through model training; Dplabtools \citep{DPLabPaper, DPLabDocs}, which offers flexible multi-magnification patch extraction; PathEX \citep{PathEX2024}, which focuses on guided patch selection; and SPLICE \citep{Alsaafin2024SPLICE}, which introduces sequential lattice-based patching for more efficient WSI coverage. OpenSlide \citep{goode2013openslide, OpenSlideSite, OpenSlideGit} provides the foundational vendor-neutral WSI reading layer that many of these tools depend on.

Despite this rich ecosystem, a gap remains: existing methods either offer fast but brittle heuristic tissue detection, or robust but computationally expensive deep learning--based detection at the patch level. No current method combines a single-pass, thumbnail-level deep learning tissue detector trained on a large heterogeneous corpus with a fully parallelized end-to-end patch extraction and embedding pipeline. AtlasPatch addresses this gap by coupling a parameter-efficiently finetuned SAM2 model with high-throughput coordinate extraction and optional GPU-accelerated embedding, achieving both robustness and scalability.

\section{AtlasPatch: Proposed Framework}\label{sec:method}

This section presents AtlasPatch, our proposed framework for scalable tissue detection and patch extraction in large-scale computational pathology. We first provide a high-level overview of the pipeline, then describe the dataset we curated along with the annotation workflow, the parameter-efficient fine-tuning of SAM2 for tissue detection, and the patch extraction and embedding components.

\subsection{Overview of AtlasPatch}\label{sec:method_overview}

AtlasPatch is a high-throughput slide-preprocessing framework designed for scalable tissue detection and patch extraction in computational pathology workflows. The pipeline operates in four sequential but optionally decoupled stages: (i)~thumbnail-based tissue detection via a finetuned SAM2 model, (ii)~patch coordinate extraction by extrapolating tissue contours to full-resolution slide coordinates, (iii)~patch embedding using widely used general-purpose and pathology-specific image encoders, and (iv)~optional patch image export. To facilitate the use of the proposed framework, we provide an accompanying command-line tool\footnote{Tool available at: \url{https://github.com/AtlasAnalyticsLab/AtlasPatch}} that can execute the full pipeline end-to-end, while also allowing each stage to be invoked independently so that users can obtain only tissue masks, patch coordinates, embeddings, or exported patch images as needed.

Fig.~\ref{fig:overview} illustrates the overall pipeline. Following the acquisition of diagnostic WSIs as multi-resolution pyramids, AtlasPatch utilizes the thumbnail for tissue detection using the finetuned SAM2 model, then extrapolates the resulting contour into the desired high resolution. This is followed by computing patch coordinates for downstream patch image export or feature embeddings. All intermediate products (masks, coordinates, features and images) are written in a consistent, slide-centric layout (HDF5 for coordinates and embeddings, PNG for optional patch images), allowing stages to be resumed or recombined. The implementation emphasizes streaming and parallel processing, in which slides are processed in parallel worker threads, and heavy operations (model inference and image embedding) are offloaded to GPUs.

\subsection{WSI dataset and annotation pipeline}\label{sec:method_dataset}

\begin{figure*}[pos=!htbp,width=\textwidth]
    \centering
    \includegraphics[width=\textwidth]{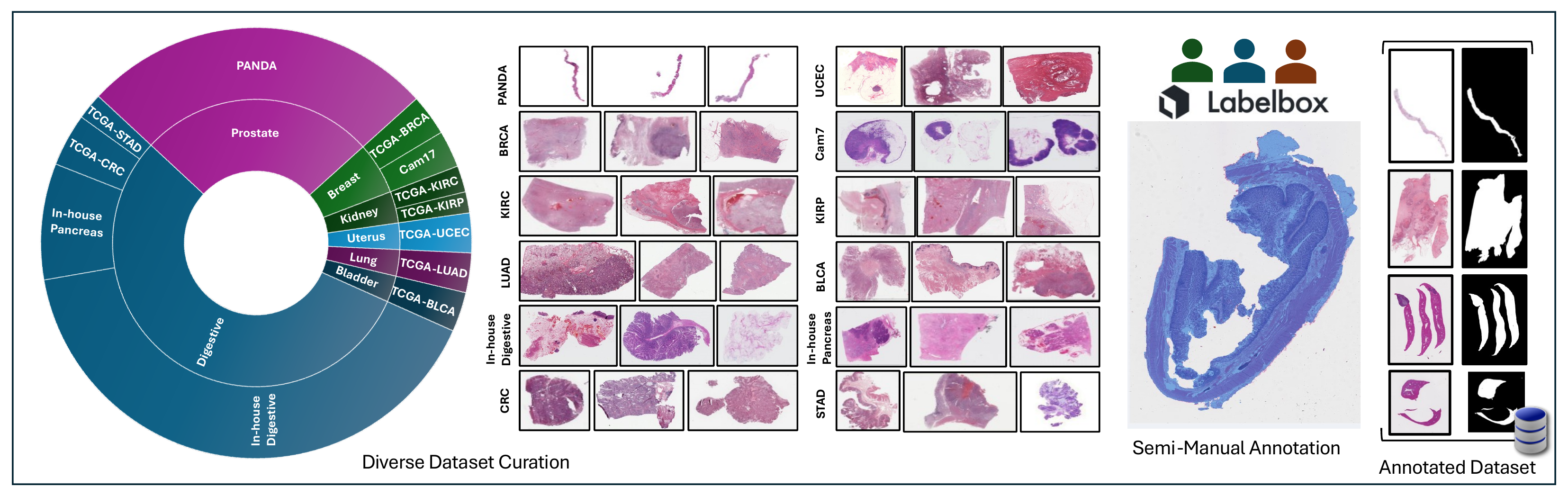}
    \caption{Diverse dataset curation and semi-manual annotation for training the tissue detector. The model in AtlasPatch is finetuned on a heterogeneous multi-organ corpus combining multiple public cohorts as well as private in-house WSIs. Slides from each dataset are curated, and annotators performed semi-manual boundary tracing of tissue on WSI thumbnails under a standardized protocol using Labelbox, yielding a large annotated dataset ($\sim$30k pairs) of tissue-versus-background masks.}
    \label{fig:dataset_summary}
\end{figure*}

To train the tissue detection component in AtlasPatch, we assembled a multi-cohort corpus of WSIs comprising 35,827 slides from 11 datasets. Two large in-house cohorts were collected from the Centre hospitalier de l'Universit\'e de Montr\'eal (CHUM) in QC, Canada (institutional review board approved project ``2025-12351''), including 3,151 pancreas WSIs and 15,736 digestive-system WSIs scanned as part of routine clinical practice. Public data were obtained from the Camelyon17 challenge \citep{bandi2018detection} for breast cancer metastases in WSIs of histological lymph node sections (1,000 slides), the PANDA prostate cancer challenge \citep{bulten2022artificial} (10,616 slides), and eight projects from The Cancer Genome Atlas (TCGA) \citep{weinstein2013cancer}, namely BLCA, BRCA, CRC, KIRC, KIRP, LUAD, STAD and UCEC, covering bladder, breast, colorectal, kidney, lung, stomach and uterine carcinomas (5,324 slides in total). Across cohorts, slides were digitized as multi-resolution pyramids at a maximum optical magnification of 20$\times$ or 40$\times$, with heterogeneous scanning hardware and acquisition settings, reflecting real-world variability in clinical workflows. For the in-house CHUM cohorts, diagnostic WSIs and associated metadata were retrieved from the institutional digital pathology archive under appropriate research agreements, de-identified before export, and screened to exclude corrupted files. Public WSIs were downloaded from their official portals and analyzed for the removal of any corrupt files. The corpus is predominantly H\&E-stained, and includes a small subset of $\sim$150 IHC-stained slides from the in-house CHUM cohort, retained to broaden stain diversity. Fig. \ref{fig:dataset_summary} summarizes the dataset creation workflow with some samples from each of the different sources.

\begin{figure}[pos=!htbp,width=\columnwidth]
    \centering
    \includegraphics[width=\columnwidth]{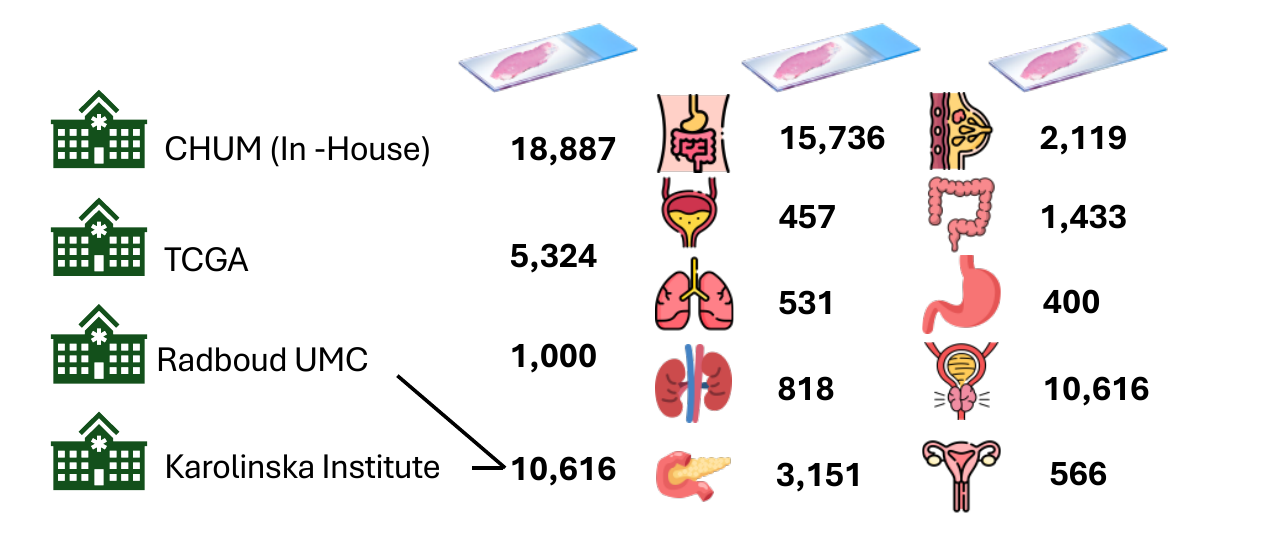}
    \caption{Composition of the $\sim$36,000 WSI thumbnail corpus across four cohorts (CHUM in-house, TCGA, Radboud UMC and Karolinska), illustrating multi-institutional and multi-organ coverage.}
    \label{fig:dataset_cohorts}
\end{figure}

\begin{figure}[pos=!htbp,width=\columnwidth]
    \centering
    \includegraphics[width=\columnwidth]{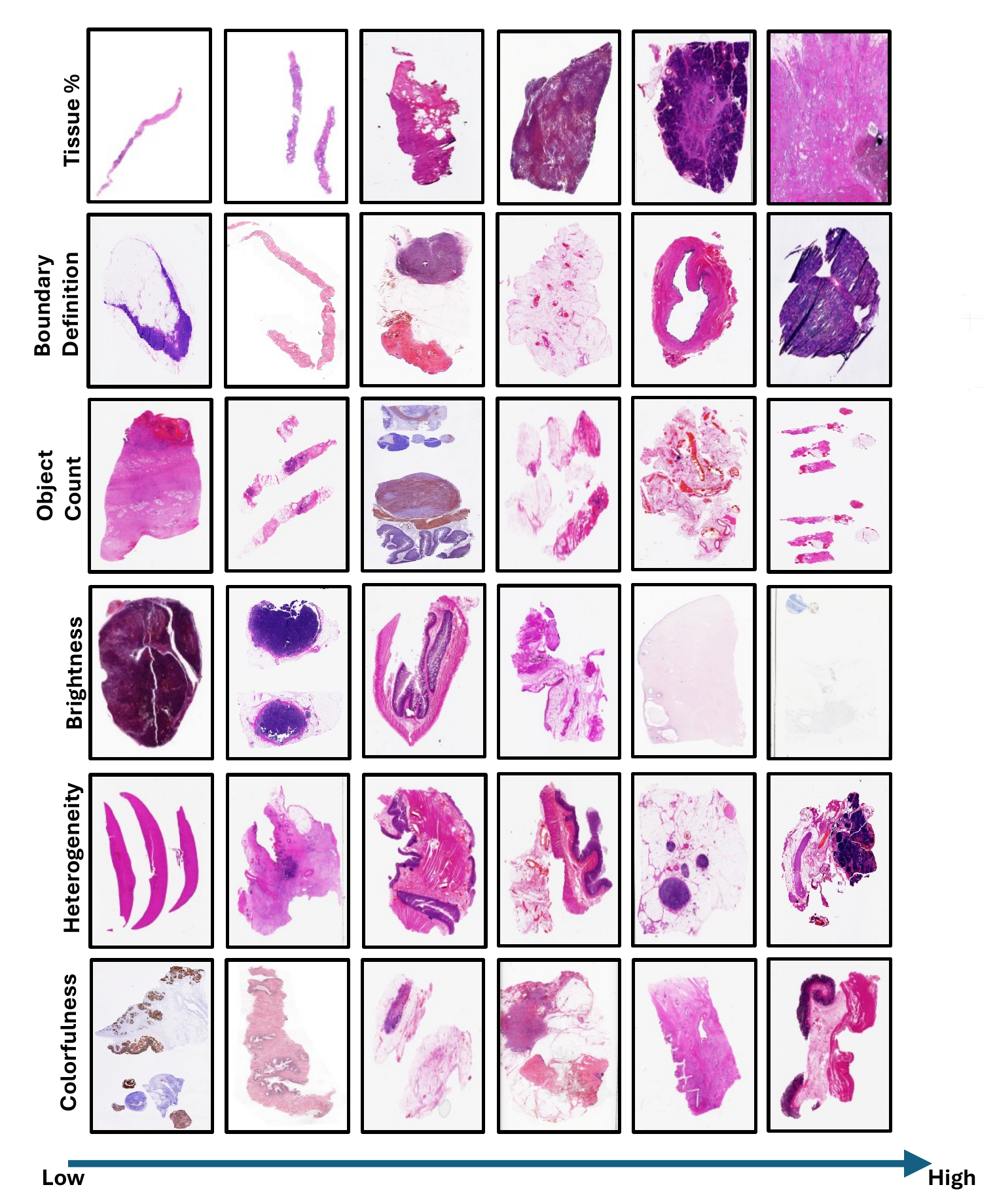}
    \caption{Example thumbnails ordered along key axes of variation, highlighting challenging edge cases for tissue detection. Rows illustrate variation in tissue percentage, boundary definition, object count, brightness, heterogeneity, and colorfulness, arranged from low (left) to high (right).}
    \label{fig:dataset_variability}
\end{figure}

To construct the thumbnail dataset used for tissue detection, WSIs were read using OpenSlide-compatible \citep{goode2013openslide, OpenSlideSite, OpenSlideGit} libraries and we extracted, for each slide, the coarsest non-empty pyramid level provided by the scanner. Thumbnails were stored as high-quality image files together with metadata indicating cohort, organ system, scanner vendor and maximum magnification. No stain normalization or color augmentation was applied at this stage, so that the subsequent segmentation model would be exposed to the natural variability of laboratory protocols and scanners. Additionally, this large corpus draws samples from four centers (in-house from CHUM, TCGA, Radboud UMC and Karolinska Institute) and spans multiple organ systems (Fig.~\ref{fig:dataset_cohorts}). This multi-institutional composition captures diversity in laboratory protocols, scanners, and acquisition settings. We also ensured that the thumbnails cover broad tissue variations (Fig.~\ref{fig:dataset_variability}) in terms of aspects like tissue coverage (percentage of the slide covered by tissue), boundary definition (contrast across the tissue boundary), object count (fragmentation), and global appearance descriptors on tissue pixels (mean brightness, hue entropy/heterogeneity, and colorfulness).

\begin{figure*}[pos=!htbp,width=\textwidth]
    \centering
    \includegraphics[width=\textwidth]{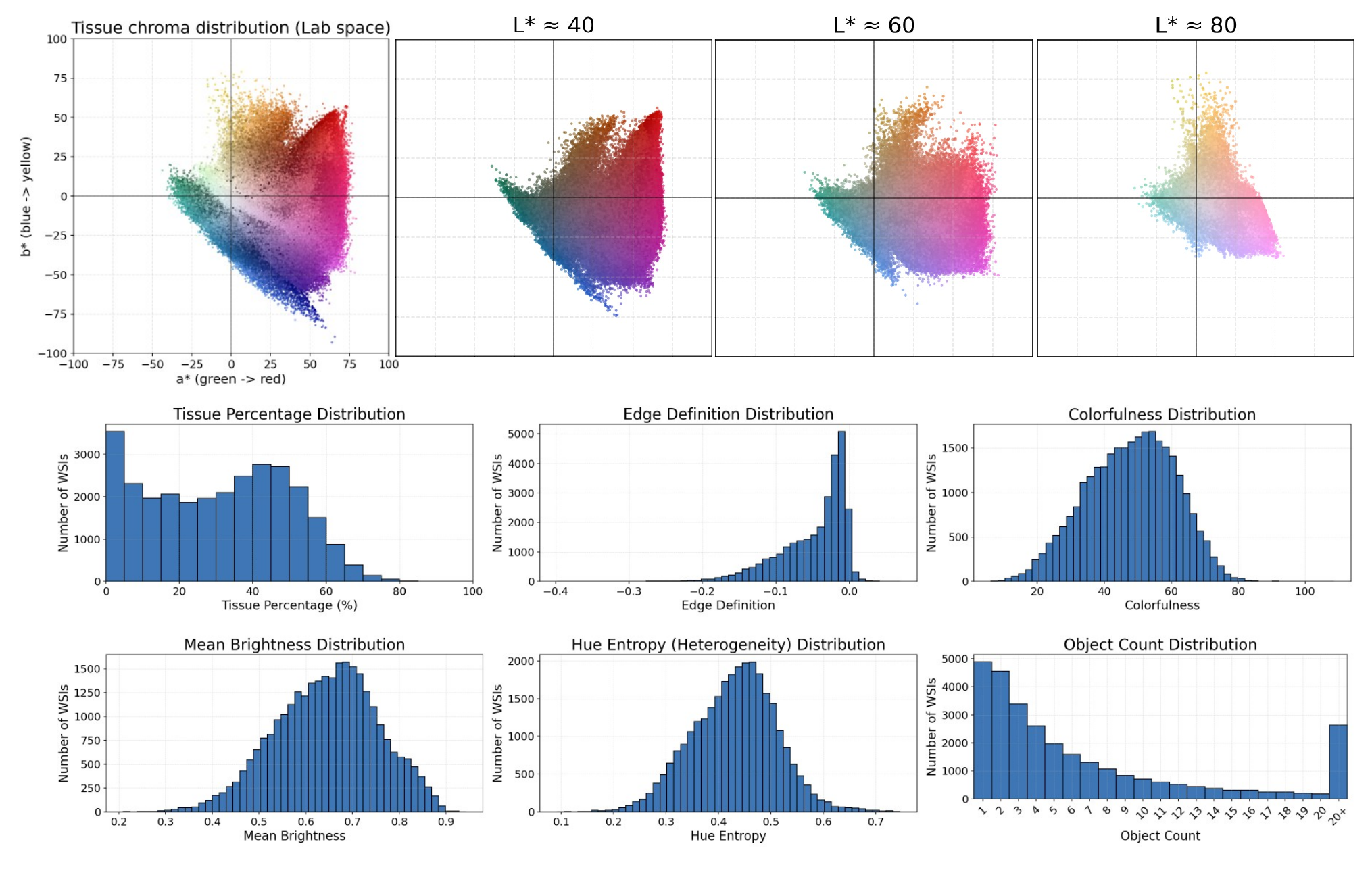}
    \caption{Quantitative characterization of the dataset variability via Lab chroma maps and slide-level statistics. The tissue chroma distribution in Lab space shows tissue pixels covering a wide region in the a*--b* plane across different lightness (L*) strata. Histograms of tissue percentage, edge definition, colorfulness, mean brightness, hue entropy, and object count confirm broad distributions consistent with real-world WSI variability.}
    \label{fig:dataset_stats}
\end{figure*}

This diversity has been quantified using slide-level statistics derived for the aforementioned attributes (Fig.~\ref{fig:dataset_stats}). From the tissue chroma distribution, we notice that tissue pixels cover a wide region in the a*--b* plane, with a* skewed strongly positive (red/magenta) while b* still spans both blue/purple (b*$<$0) and yellow/brown (b*$>$0) components. This pattern is consistent across lightness (L*) strata, although the spread varies with illumination level. This broad but structured spread mirrors the staining and acquisition variability encountered in routine practice, supporting the diversity of our slide collection. The histograms further confirm this real-life variance across tissue geometry and appearance. Tissue percentage exhibits a wide spread, with many low-coverage slides (sparse biopsies/small fragments) and progressively fewer high-coverage cases (large resections). Object count is strongly long-tailed, indicating that while many thumbnails contain only a few connected tissue components, a non-trivial subset is highly fragmented into many islands. Edge definition (signed Michelson contrast) is concentrated near zero with a long negative tail, where negative indicates brighter background than tissue, values near zero reflect weak boundary contrast, and larger-magnitude negatives correspond to cleaner tissue--background separation. Finally, brightness, hue entropy (heterogeneity), and colorfulness show broad and largely unimodal distributions, consistent with the variability in stain intensity, color complexity, and acquisition conditions encountered in routine WSI practice, supporting the diversity of the slide cohort.

\begin{figure*}[pos=!htbp,width=\textwidth]
    \centering
    \includegraphics[width=\textwidth]{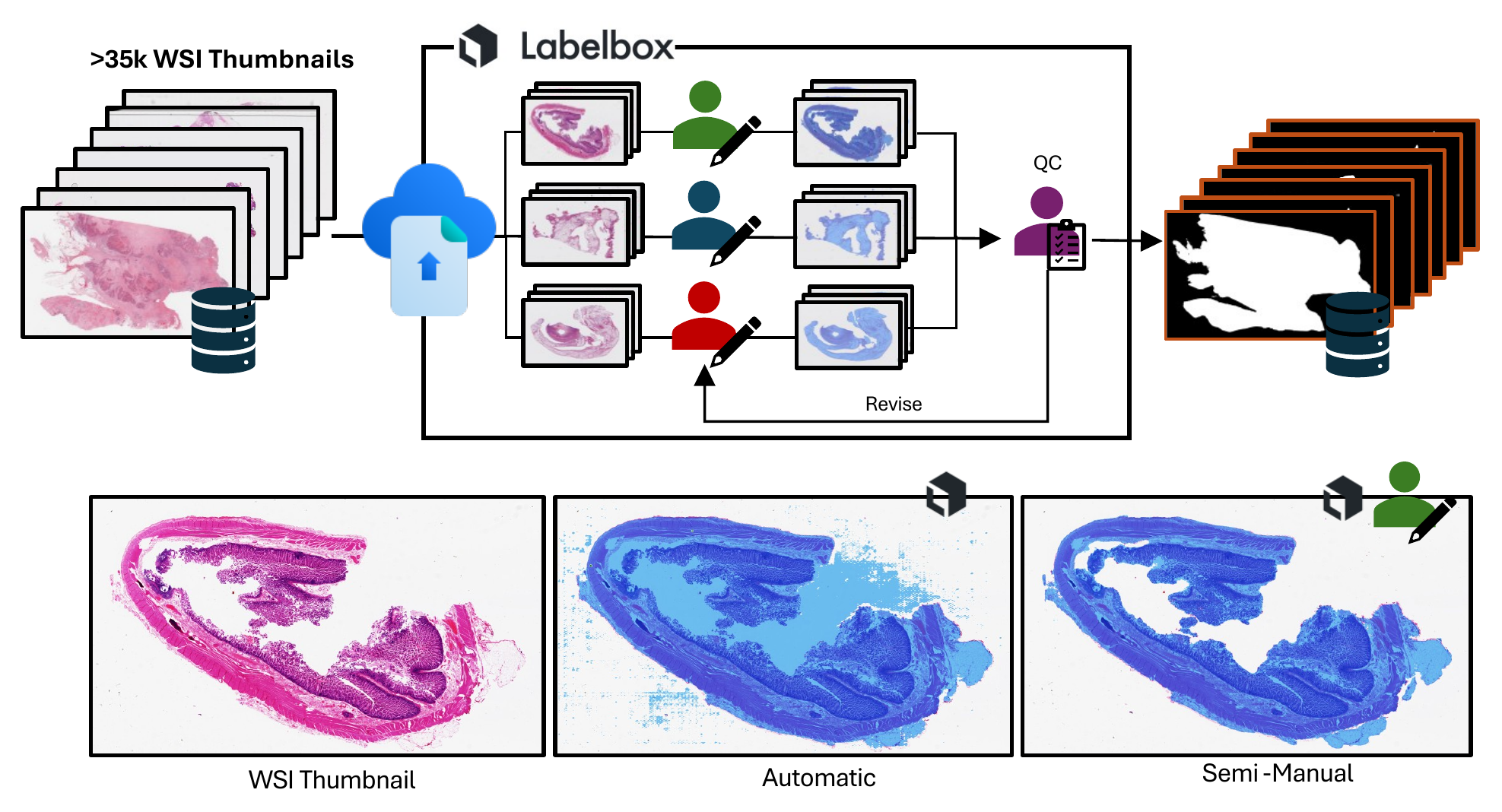}
    \caption{Semi-automatic annotation pipeline in Labelbox. Over 35,000 WSI thumbnails are uploaded and distributed to annotators, who refine pre-segmentation with manual corrections. A quality controller (QC) reviews each mask and returns ambiguous cases for revision, yielding high-quality tissue-versus-background masks for SAM2 finetuning and evaluation. The bottom row shows an example of a WSI thumbnail, its automatic segmentation (often insufficient), and the refined semi-manual annotation.}
    \label{fig:annotation_pipeline}
\end{figure*}

To obtain reliable masks for training the tissue detector, we created ground-truth tissue-versus-background masks on these thumbnails via a structured semi-manual workflow in Labelbox \citep{labelbox}. The collected dataset was split on three annotators, who initially underwent several training rounds on pilot slides, during which their annotations were systematically inspected by a quality-control (QC) reviewer and aligned through feedback to ensure consistent interpretation of the guidelines. Within Labelbox, annotators primarily relied on the AutoSegmentBox tool to identify tissue regions by drawing bounding boxes at multiple scales to capture both large tissue areas and finer structures. The resulting automatic masks were then carefully refined with the Pen tool to adjust boundaries, recover small tissue islands, and remove artifacts. By combining AutoSegmentBox at different zoom levels with targeted Pen-based labeling, annotators efficiently produced detailed tissue masks. An initial pass of annotations was followed by a QC step in which a more senior reviewer inspected each mask, requesting corrections in cases of under- or over-segmentation or inconsistent inclusion of artifacts (Fig.~\ref{fig:annotation_pipeline}). This process yielded approximately 30,000 high-quality thumbnail--mask pairs spanning varying organs, cohorts, and acquisition conditions. Thumbnails and annotations were vetted by a board-certified pathologist. The resulting corpus provides dense tissue-background supervision designed to support robust, cross-institution generalization. Altogether, this highly variable and carefully annotated dataset is central to the design of a slide preprocessing tool that is robust to the diverse slide appearances encountered in downstream pipelines.

To quantify the diversity of the corpus, we computed a set of slide-level statistics, as reported in Fig.~\ref{fig:dataset_stats}. Using the binary tissue mask, we measured tissue coverage (defined as the fraction of thumbnail pixels labeled as tissue), object count (defined as the number of connected tissue components after a mild morphological cleanup to remove tiny specks), and boundary definition, a contrast-based index computed as the Michelson contrast between mean grayscale intensities just inside and just outside the tissue boundary, with higher magnitudes indicating a sharper separation between tissue and background. On the RGB thumbnail restricted to tissue pixels, we derived global appearance descriptors including mean brightness, computed by converting the thumbnail to grayscale, averaging pixel values within the tissue mask and normalizing by 255 to obtain a 0--1 score. We also computed heterogeneity, quantified as the Shannon entropy of the tissue hue histogram, and colorfulness, using the Hasler--S\"usstrunk opponent-color metric on tissue pixels. To further characterize color variability, we additionally constructed a Lab chroma map by randomly sampling 1\% of tissue pixels from each thumbnail in the full dataset, converting their RGB values to CIE Lab, and plotting their distribution in the a*b* plane for different lightness levels. For each scalar statistic, we computed slide-level histograms over the full corpus. These histograms were also used to define low/medium/high bins when constructing the stratified subsets in the data-diversity experiments (Fig.~\ref{Fig5}).

For training the tissue detection backbone (SAM2), we applied an 80:10:10 split on the full set of $\sim$30,000 annotated thumbnails, yielding training, validation and internal test subsets for tissue detection. For the downstream MIL benchmarks, each MIL method was evaluated using 10-fold cross-validation on each slide-level dataset.

\subsection{Parameter-efficient SAM2 adaptation for tissue detection}\label{sec:method_sam2}

The Segment Anything Model 2 (SAM2) \citep{ravi2024sam} is a vision foundation model designed for prompt-based image and video segmentation. Building upon the original SAM \citep{Kirillov2023SAM}, SAM2 introduces a fully trainable vision transformer architecture equipped with streaming memory for real-time and temporally consistent segmentation. It employs a modular encoder--decoder design with a hierarchical vision transformer backbone that efficiently processes multi-scale features while maintaining a global receptive field. The architecture consists of four main components: (i)~an image encoder, (ii)~a prompt encoder, (iii)~a mask decoder, and (iv)~a memory module comprising memory attention and a memory bank. The image encoder projects each frame into high-dimensional patch embeddings and extracts contextual features across multiple scales. The prompt encoder embeds user inputs, such as points, bounding boxes, or coarse masks, into a latent space aligned with the image representations. The mask decoder fuses these image and prompt embeddings via cross-attention to produce one or multiple segmentation masks, each associated with a confidence score.

AtlasPatch operates on independent WSI thumbnails, and therefore uses SAM2 in an image-only setting without leveraging the memory bank or temporal propagation. We nonetheless adopt SAM2 (rather than the original SAM) because it is pretrained at larger scale and provides 6$\times$ faster inference than SAM due to its more efficient image encoder backbone, being the Hiera hierarchical backbone instead of a ViT-based encoder, making it better suited for high-throughput thumbnail segmentation in large WSI cohorts. SAM2's Hiera backbone is available in different sizes being tiny, small, base, and large.

\begin{figure*}[pos=!htbp,width=\textwidth]
    \centering
    \includegraphics[width=\textwidth]{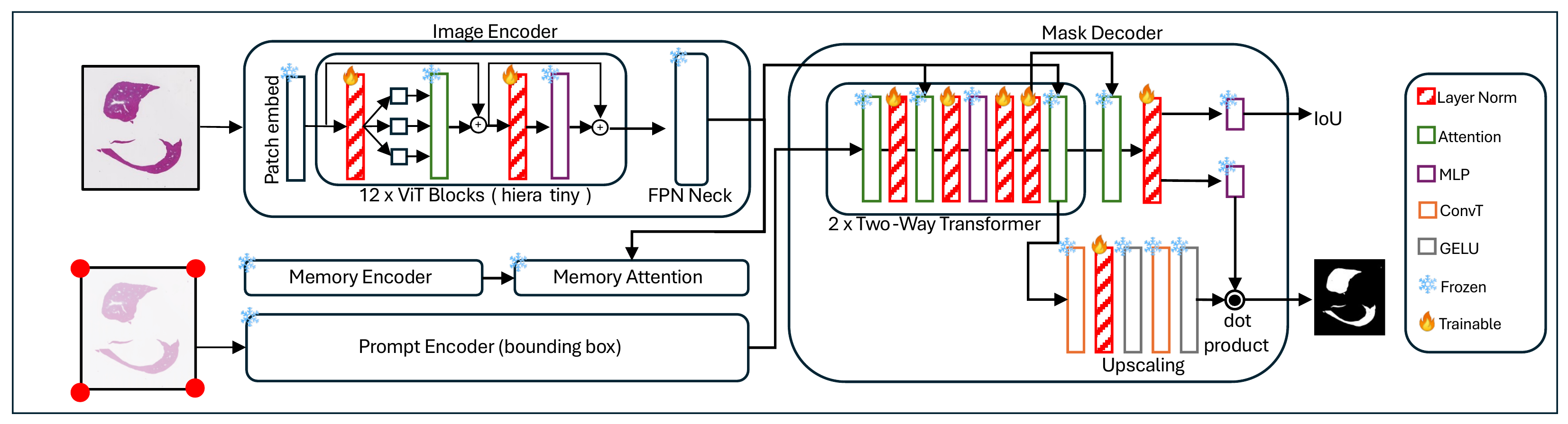}
    \caption{Thumbnail-based efficient finetuning of SAM2. The segmentation model is trained on the curated thumbnails dataset by only finetuning the normalization layers of the SAM2 model, which represent approximately 0.076\% of the parameters in the hiera-tiny variant of SAM2. The image encoder backbone and all attention/MLP weights remain frozen (blue snowflake), while only the Layer Normalization affine parameters are updated (orange flame).}
    \label{fig:sam2_finetuning}
\end{figure*}

While SAM2 demonstrates strong generalization on natural images and even some medical imaging modalities, directly applying it to histopathology images is non-trivial. The adaptation of SAM2 to histopathological images presents challenges due to the distribution shift between natural and microscopic imagery. SAM2 is pretrained on billions of natural images where structures, colors, and object boundaries differ dramatically from histological tissue patterns. In WSIs, tissue and background boundaries can be subtle, textures are stochastic rather than object-like, and color distributions are dominated by H\&E staining. Consequently, the pretrained features may fail to align with histopathological semantics unless adapted with domain-specific fine-tuning.
With hundreds of millions of parameters, fully fine-tuning foundation models such as SAM2 is impractical. The relatively small scale of annotated histopathology datasets increases the risk of overfitting and catastrophic forgetting of the pretrained representations. Moreover, full parameter updates demand substantial computational resources and memory, making full fine-tuning both inefficient and costly for large models. These limitations have motivated the development of lightweight and targeted adaptation techniques referred to as Parameter-Efficient Fine-Tuning (PEFT), which aim to achieve comparable performance to full fine-tuning while updating only a small fraction of model parameters. Inspired by recent parameter-efficient fine-tuning strategies \citet{zhaotuning, rostami2025segmentcrackdeepsemantic}, we adapt SAM2 through \emph{Layer Normalization Fine-Tuning} (Fig. \ref{fig:sam2_finetuning}), which selectively updates only the affine parameters of normalization layers, which are given as:

\begin{equation}
\text{LN}(\mathbf{x}) = \gamma \frac{\mathbf{x}-\mu(\mathbf{x})}{\sigma(\mathbf{x})+\epsilon} + \beta,
\label{eq:layernorm}
\end{equation}
where $\mathbf{x}$ denotes the input to an LN layer, $\mu(\mathbf{x})$ and $\sigma(\mathbf{x})$ are the per-feature mean and standard deviation, $\gamma$ and $\beta$ are learnable parameters, and $\epsilon$ is a small constant for numerical stability. During layer normalization fine-tuning, only $\gamma$ and $\beta$ are updated via backpropagation, effectively re-scaling and re-centering the pretrained feature activations to match the target WSI domain.

The rationale behind the method is that Layer Normalization (LN) plays a crucial role in transformer architectures by stabilizing feature distributions across tokens and enabling robust gradient propagation. In the context of domain adaptation, the affine parameters of LN layers (scale $\gamma$ and shift $\beta$) capture domain-specific feature statistics without altering the global representational structure. Thus, adapting only these small subsets of parameters can yield substantial domain alignment at minimal computational cost. The method has shown superior performance on the original SAM compared with alternative PEFT approaches such as LoRA and adapters, demonstrating strong adaptation with minimal parameter updates \citep{rostami2025segmentcrackdeepsemantic}. Building on these findings, the approach was used in this work for SAM2 to achieve efficient and stable domain adaptation.

In this work, the \emph{SAM2-Tiny} variant was employed, which consists of approximately 40~million parameters, initialized with publicly released pretrained weights. The Hiera-tiny variant proved sufficient for the tissue detection task given our efficient finetuning, as shown in Fig.~\ref{Fig6}. Fine-tuning was performed on the curated WSI thumbnail--mask pairs. Each input image was resized to $1024\times1024$ pixels and paired with its corresponding binary mask. During training, a bounding-box prompt was automatically generated to encompass the entire image region, ensuring full spatial coverage and allowing the model to learn tissue segmentation without relying on manual or localized prompts. This prompt type provides a simple yet effective supervision signal, enabling SAM2's prompt encoder to encode the global spatial context while allowing the mask decoder to delineate fine-grained tissue boundaries.

The fine-tuning was conducted using mixed-precision training with a batch size of 2, a learning rate of $5\times10^{-4}$, and the AdamW optimizer coupled with a cosine-annealing learning rate scheduler to enable stable convergence. The loss function was a weighted combination of Dice loss ($L_\mathrm{Dice}$) and binary cross-entropy loss ($L_\mathrm{BCE}$), shown in \citet{rostami2025segmentcrackdeepsemantic} to effectively balance region overlap accuracy with pixel-wise consistency:
\begin{equation}
L = 0.65 \, L_\mathrm{Dice} + L_\mathrm{BCE}.
\label{eq:loss_combined}
\end{equation}
Training was conducted for 50 epochs with early stopping based on the validation F1-score to prevent overfitting. In inference, the model processes each WSI thumbnail once using a full-image bounding-box prompt to predict the tissue mask. This thumbnail-level segmentation eliminates the need for exhaustive tiling and enables fast and memory-efficient ROI detection with negligible computational overhead.


\subsection{Patch coordinate extraction}\label{sec:method_patching}

For tissue detection, WSIs are first converted to thumbnails at a target objective power (default $\times$1.25), using the scanner's native pyramid levels and optional downscaling. When processing multiple slides, thumbnail preparation is parallelized across a pool of worker threads so that I/O and image resizing overlap. The resulting thumbnails are then passed in batches to our fine-tuned SAM2 model, which runs batched segmentation and returns binary tissue-versus-background masks. Masks are stored and can be visualized as overlays on the original thumbnails for quality control.

For tissue patch coordinate extraction, AtlasPatch converts each tissue mask to polygonal contours with explicit handling of holes and very small fragments, discarding tissue regions below a configurable area threshold. The contours are mapped from thumbnail space to level-0 slide coordinates, and the tool automatically selects an appropriate pyramid level and read window size to match the requested target magnification. A regular grid is then walked over each contour's bounding box with a configurable step size, and candidate patch locations are accepted only if a central point or one of the probes (corners) lie within tissue and outside holes. In its default mode, the extractor yields only patch coordinates (without reading pixels) and writes them to per-slide HDF5 files along with metadata describing level-0 patch footprint, magnification and overlap. When multiple slides are processed, each slide's extraction runs in its own worker thread, with a global cap on concurrently open WSIs for proper memory management and per-slide lock files to prevent duplicate work across processes.

\subsection{Embedding extraction and patch export}\label{sec:method_embedding}

For patch embedding and image export, AtlasPatch reopens the HDF5 coordinate files, iterates over stored patch locations and re-reads the corresponding regions from the original WSIs at the predetermined size. Patch embedding is organized per extractor and per slide: for each chosen encoder from the internal registry (including convolutional and transformer-based general-purpose and pathology-specific models, as well as user-defined plugins). Per-slide lock files ensure that concurrent runs never corrupt feature groups and allow caching of existing embeddings to skip redundant computation. When the optional image-saving flag is enabled, RGB patches are dispatched to a bounded thread pool for PNG encoding and disk writes while coordinate iteration continues, overlapping CPU-bound image I/O with ongoing slide processing.

Taken together, this modular design makes AtlasPatch adaptable to a wide range of users and workflows. Clinically oriented teams can rely on it purely as a fast and robust tissue detector, while method developers can plug in their own encoders on top of the available standardized encoders. Large-scale studies can leverage its batch processing, device configuration and parallel I/O to run efficiently on workstations or HPC clusters. Because each stage is independently configurable and can be stopped or resumed, the tool can be easily tailored to new datasets and downstream tasks without changing the core implementation.

\section{Experiments and Results}\label{sec:experiments}
This section evaluates AtlasPatch in terms of tissue detection performance, robustness across heterogeneous WSI conditions, computational efficiency, and downstream performance in slide-level classification tasks. 

\subsection{Experimental setup and baselines}
This section presents the experimental setup used to evaluate AtlasPatch, including the baseline methods used for comparison and the downstream slide-level classification tasks.

\subsubsection{Baselines and implementation details}\label{sec:method_baselines}

For tissue detection, we benchmarked widely used open-source baselines that span (i)~classical thumbnail thresholding, and (ii)~AI-based tissue detectors that operate on thumbnails or WSI tiles. For the thresholding methods, we evaluated the tissue detector shipped with CLAM \citep{CLAMGit, Lu2021CLAM}, the tissue masking utilities in TIAToolbox \citep{pocock2022tiatoolbox, TIADocs}, and other commonly used pipelines including HistoQC \citep{Janowczyk2019HistoQC, HistoQCGit}, dplabtools \citep{DPLabDocs, DPLabPaper}, and EntropyMasker \citep{EntropyMasker, EntropyMaskerGit}. HistoQC identifies tissue by thresholding a low-resolution slide thumbnail for bright/dark content and color variance, then refines the mask using morphological operations to suppress background, adipose-like regions, and small artifacts. Dplabtools operates on a user-specified slide magnification, generating a binary tissue mask via thresholding in HSV/Lab space (or Otsu), with optional hole filling/dilation and pruning of small objects. TIAToolbox provides Otsu- and morphology-based TissueMasker utilities that build tissue-versus-background masks from thumbnails, and integrates these masks directly into its patch extraction utilities so patches can be filtered based on tissue regions using either an automatically generated or user-supplied mask. Finally, although CLAM is primarily an end-to-end MIL pipeline, its patch extraction component is frequently used as a standalone method. The tissue detection component in this method converts the slide to HSV, applies median filtering and thresholding (optionally Otsu), and extracts tissue contours while removing holes and small components via area-based contour filtering. For all these baselines, we used their corresponding public repositories and retained default parameters whenever configurable options were available to avoid cohort-specific tuning. Because not every baseline produces binary tissue mask exports as its final output (e.g., some return contours, probability maps, or patch-wise outputs), we added minimal adapters where needed so that all methods could be evaluated under a common binary mask representation. For AI-based tissue detection, we benchmarked with SAM2 as a ``plug-and-play'' foundation model baseline to test how far a strong generic segmenter can go without any domain adaptation. We applied SAM2 directly to WSI thumbnails and converted its predicted regions into tissue-versus-background masks. The TRIDENT \citep{Zhang2025Trident, TRIDENTGit, TRIDENTDocs} WSI processing tool provides several tissue detectors. We benchmarked against the GrandQC \citep{weng2024grandqc} and Hest \citep{jaume2024hest} tissue detection variants in that tool. Both variants operate at the patch-level, where the model is fed tissue patch images and outputs the corresponding masks, which are later stitched to form slide-level mask. GrandQC operates on patches extracted from the WSI thumbnail, while Hest operates on patches extracted from higher magnification (10$\times$).

For end-to-end preprocessing pipelines (tissue detection through patch coordinate extraction), we focused on tools that support WSI loading, tissue detection, and coordinate export for downstream feature encoding and MIL. For both CLAM and Trident tools, at the user-specified magnification/patch size, patch extraction slides a window over each contour's bounding box at the chosen pyramid level, retains coordinates whose patch footprint lies within the tissue region, passes simple white/black heuristics for further filtering, and writes accepted patch coordinates into per-slide HDF5 ``bag'' files. Other tools considered in the tissue-detection benchmark were not included in the full-pipeline comparison because they either stop at tissue masking (i.e., do not provide a complete coordinate export/patchification pipeline), or are older/out-of-date in ways that complicate large-scale reproducible WSI fetching and processing across heterogeneous cohorts.

To ensure fair and reproducible comparisons, we standardized both compute settings and evaluation protocol across baselines. For AI-based tissue detection models, inference was executed on the same computational infrastructure using a single RTX Ada 6000 GPU with a fixed batch size of 32 and 4 workers, and segmentation performance was reported on a held-out test set of 3,000 slides. Whenever thresholding tools allow for parallel workers, we use 4 workers. For computational complexity analysis, covering tissue segmentation runtime and, where applicable, full end-to-end preprocessing runtime, we again used the same infrastructure for all tools (RTX Ada 6000) and constrained each method to one GPU when GPU execution was supported. We further fixed downstream patch extraction-related parameters across pipelines to 512$\times$512 patch size, 0 overlap, 20$\times$ magnification, batch size 32, and 4 workers. Runtime comparisons were conducted on a shared subset by sampling 100 slides from the cohort and running every method on the exact same slide set, ensuring that both segmentation and pipeline-level complexity reflected differences in implementation rather than differences in data, hardware, or configuration.

\subsubsection{Downstream classification tasks}
For the downstream evaluation, we considered six slide-level prediction tasks spanning both in-house and public cohorts, where multiple-instance learning (MIL) was used on bags of tissue patches to aggregate patch-level representations to predict slide label. For each task, WSIs were first processed by a given preprocessing tool to obtain tissue patch coordinates and embeddings, and these bags were then passed to several MIL methods (ABMIL \citep{abmil}, CLAM \citep{Lu2021CLAM}, DFTD \citep{dtfd}, DSMIL \citep{dsmil}, MeanMIL \citep{shaomultiple}, RRT \citep{rrt}, TransMIL \citep{transmil}, and WIKG \citep{wikg}). The performance reported in Fig.~\ref{Fig7} is averaged over MIL methods for each tool--dataset pair. All patch images were embedded using the UNIv1 \citep{chen2024towards} foundation model.

Two tasks were derived from an in-house digestive pathology cohort of colorectal biopsies and resections. The Invasiveness task defines a binary label indicating the presence versus absence of invasive colorectal carcinoma on a slide, as determined from the original pathology report. The Dysplasia task uses the same corpus but assigns a dysplasia-level label (non-dysplastic, low-grade or high-grade dysplasia) at the slide level based on routine diagnostic grading in the report.

Three tasks are based on The Cancer Genome Atlas (TCGA) cohorts. For breast carcinoma subtyping (BRCA), we used diagnostic H\&E WSIs from the TCGA-BRCA project and restricted to cases annotated as invasive ductal carcinoma (IDC) or invasive lobular carcinoma (ILC). For renal cell carcinoma subtyping (KIRCvsKIRP), we combined WSIs from TCGA-KIRC (Kidney Renal Clear Cell Carcinoma) and TCGA-KIRP (Kidney Renal Papillary Cell Carcinoma) and assigned labels accordingly. For lung carcinoma subtyping (LUADvsLUSC), we analogously pooled TCGA-LUAD and TCGA-LUSC WSIs and labeled slides as lung adenocarcinoma or lung squamous cell carcinoma.

Finally, the prostate cancer grading task uses the PANDA challenge dataset, consisting of digitized prostate biopsies. We adopted the official slide-level Gleason grade group labels provided with PANDA metadata, treating grade group prediction as a multi-class classification problem.

\subsection{Tissue detection performance}\label{sec:exp_performance}

Tissue detection performance measures how accurately a preprocessing method delineates biologically meaningful tissue from background and artifactual structures in whole-slide images. This step is fundamental because the resulting tissue mask defines the search space for downstream patch extraction: missed tissue leads to information loss, whereas over-segmentation increases background-heavy sampling and unnecessary computational cost. 

\subsubsection{Tissue detection qualitative performance}

\begin{figure*}[pos=!htbp,width=\textwidth]
    \centering
    \includegraphics[width=0.8\textwidth]{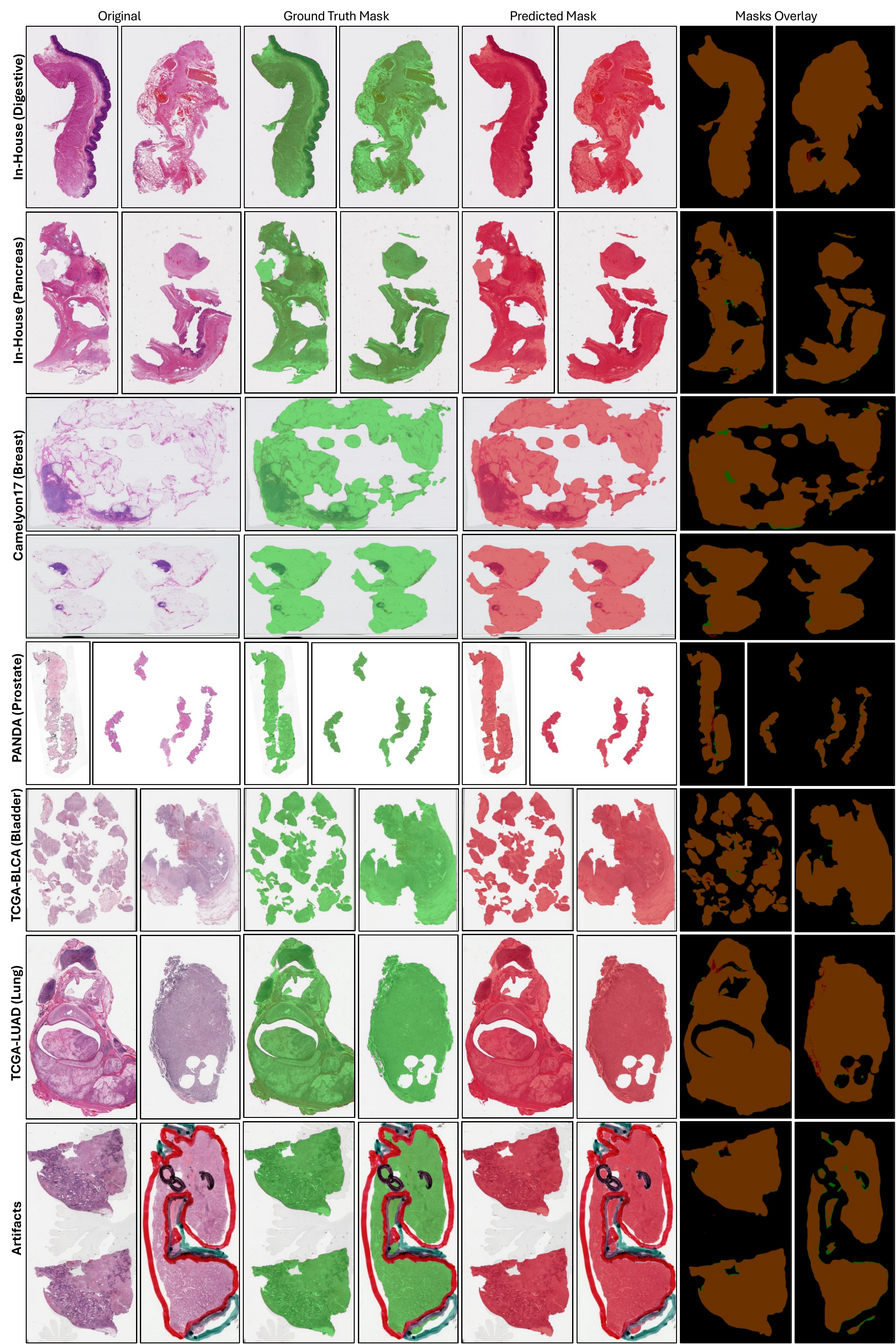}
    \caption{Qualitative examples of annotated masks as well as AtlasPatch tissue masks across datasets and artifacts. Representative WSI thumbnails from multiple sources are shown alongside their annotated ground-truth tissue masks, AtlasPatch predictions, and mask overlays. Ground-truth tissue is displayed in green, model predictions in red, and overlapping regions in brown, illustrating close agreement between AtlasPatch and annotations across organs and cohorts. The bottom row shows challenging cases with scanner or preparation artifacts, where AtlasPatch masks largely follow true tissue while ignoring non-tissue structures. These samples come from a testing set that was not seen by the model during training.}
    \label{Fig3}
\end{figure*}

Fig.~\ref{Fig3} shows representative thumbnails alongside ground-truth masks, AtlasPatch predictions, and overlays (ground truth in green, predictions in red, and overlap in brown). It can be seen that the overlays are dominated by overlap across the different cohorts and organs. Beyond cohort and organ shifts, AtlasPatch handles substantial variation in tissue geometry, accurately segmenting both large contiguous resections and highly fragmented biopsies with many small tissue islands while largely excluding background. In some cases, the model predictions even improve on the ground truth mask, as seen in the top Camelyon17 sample in Fig.~\ref{Fig3}, where the predicted mask excludes a white background region that was missed by the annotators. Furthermore, we show cases with common scanner and preparation artifacts (e.g., ink and other non-tissue structures). It can be seen that the tissue detector in AtlasPatch confines its predictions to biologically plausible tissue regions and efficiently excludes most artifactual structures.

We attribute this combination of cross-cohort generalization, robustness to diverse tissue complexities and geometries, improvements beyond annotations, and artifact suppression to two main factors: (i)~the breadth of our multi-institutional training corpus and the efficiency of the annotation process, in which masks explicitly exclude common non-tissue artifacts and span a wide range of tissue presentations, and (ii)~the efficient finetuning strategy that updates only a small subset of model parameters, coupled with the pretrained SAM2 backbone.

\begin{figure*}[pos=!htbp,width=\textwidth]
    \centering
    \includegraphics[width=0.8\textwidth]{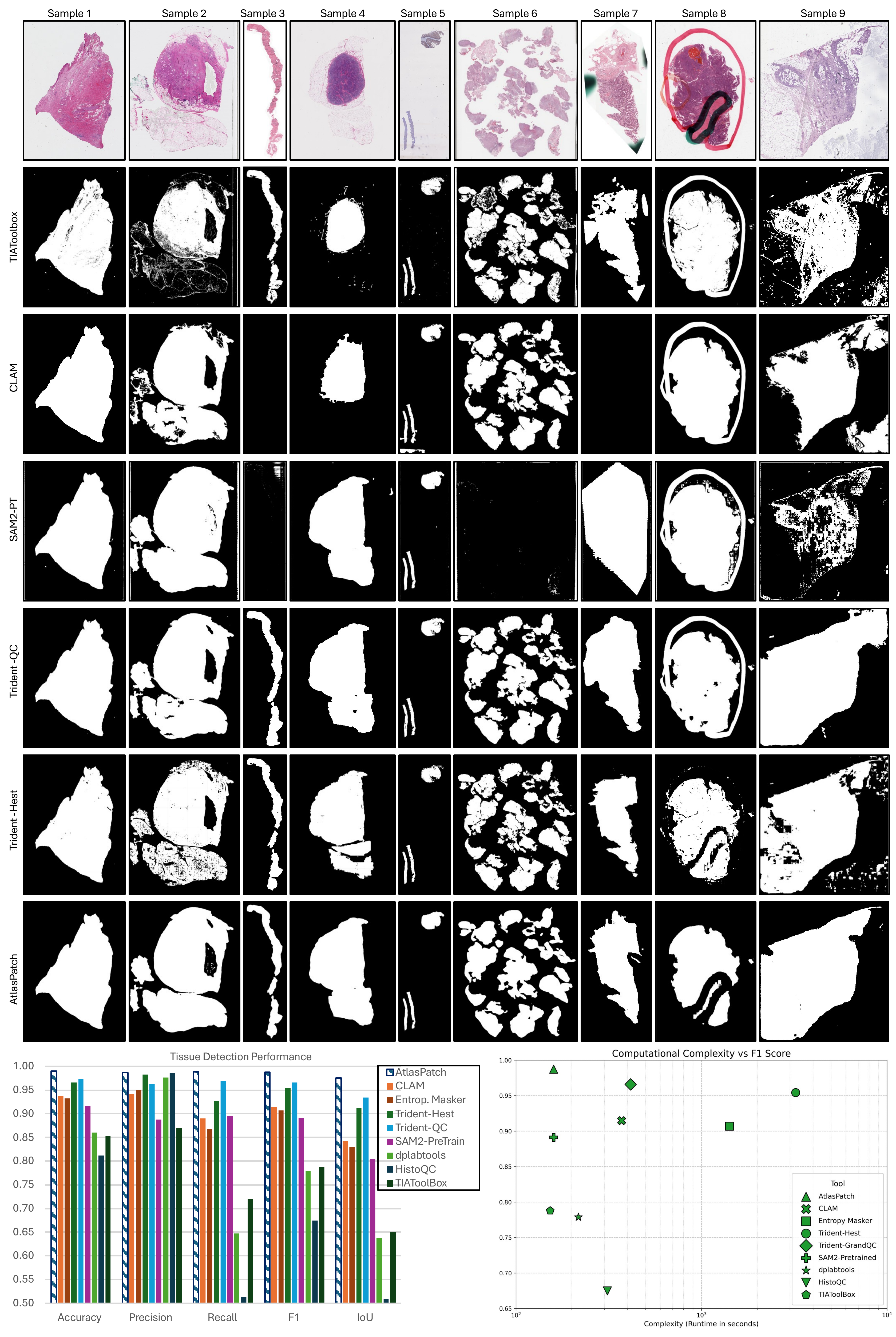}
    \caption{Quantitative and qualitative analysis of AtlasPatch tissue detection against existing slide-preprocessing tools. Representative WSI thumbnails are shown from diverse tissue features and artifact conditions, with tissue masks predicted by thresholding methods (TIAToolbox, CLAM) and deep learning methods (pretrained ``non-finetuned'' SAM2 model, Trident-GrandQC, Trident-Hest, and AtlasPatch), highlighting differences in boundary detection, artifact suppression, and handling of fragmented tissue. Tissue detection performance is also shown on the held-out test set, highlighting that AtlasPatch matches or exceeds the other methods. The segmentation complexity--performance trade-off shows F1-score against segmentation runtime (per 100 WSIs), showing AtlasPatch achieves high performance with substantially lower wall-clock time than patch-based methods, underscoring its suitability for large-scale WSI preprocessing.}
    \label{Fig4}
\end{figure*}

We evaluate the AtlasPatch tissue detector against widely used thresholding- and deep learning-based methods. Fig.~\ref{Fig4} summarizes comparisons qualitatively on representative slide thumbnails drawn from diverse organs and tissue conditions (brightness, fragmentation, sparsity, scanner artifacts, and pen markings), and quantitatively using segmentation metrics on the held-out test set and a runtime-performance analysis. Thresholding-based methods show limited robustness to the heterogeneity present in WSIs. Across multiple samples in Fig.~\ref{Fig4}, CLAM and TIAToolbox can miss substantial tissue when contrast or stain appearance deviates from typical cases, and they frequently fail to distinguish artifacts from tissue. Consistent with these qualitative observations, we found that the CLAM tool fails to detect any tissue at all in roughly half of the prostate WSIs from the PANDA dataset. The pretrained (non-finetuned) SAM2 model is also not a reliable out-of-the-box tissue detector. Without histopathology domain exposure, it inconsistently captures tissue and can include visually salient non-tissue patterns, reflecting the gap between the generic object detection in SAM2 and the more subtle cues required for robust tissue detection. Patch-based deep learning methods offer improved robustness over thresholding-based baselines. However, Trident-GrandQC frequently misinterprets artifacts as tissue, particularly in slides with strong pen markings or scanner streaks. Trident-Hest shows more resilient behavior with respect to artifacts but can miss tissue at ambiguous boundaries. While these methods operate at the patch level and are expected to recover fine-grained tissue details, they can lose global context which sometimes could be beneficial to classify ambiguous regions. In contrast, AtlasPatch delivers consistent tissue masks across all the scenarios shown. It accurately outlines tissue in low-contrast slides, preserves small tissue fragments in highly fragmented biopsies and effectively suppresses non-tissue artifacts. While AtlasPatch does not always reproduce every fine-grained internal hole within tissue regions, it systematically captures the full spatial footprint of biologically meaningful tissue and avoids large false negatives.

\subsubsection{Tissue detection quantitative performance}
Quantitatively, on the held-out test set of 3,000 WSIs, AtlasPatch matches or exceeds competing tools across accuracy, precision, recall, F1 and IoU (Fig.~\ref{Fig4}). Precision measures how much of what a method labels as tissue is truly tissue, whereas recall captures how much of the annotated tissue is successfully recovered, with F1 being their harmonic mean and IoU directly measuring the overlap between predicted and ground-truth tissue masks. These metrics are more meaningful compared to pixel accuracy, which can be inflated by the large background regions. Particularly, AtlasPatch has strong precision (0.986), slightly higher than the much more complex Trident-Hest (0.983), dplabtools (0.977), and HistoQC (0.985) as the nearest competitors. Despite outperforming competing methods across all metrics, we highlight precision because pixel-level ground truth can be ambiguous at tiny internal boundaries, and methods may correctly exclude these regions as background, which can be unfairly penalized in accuracy and recall, whereas precision primarily reflects whether predicted tissue pixels are truly tissue. The computational complexity plot further shows AtlasPatch on a favorable performance-complexity frontier. Heuristic/thresholding methods are fast but less robust, whereas patch-based deep models incur substantially higher runtime due to processing many (up to thousands) patches per slide. By operating directly at thumbnail resolution with a single SAM2-based forward pass and an efficient finetuning strategy on a diverse dataset, AtlasPatch achieves high segmentation quality with markedly lower wall-clock time than patch-wise detectors (2.6$\times$ faster than Trident-GrandQC and 20$\times$ faster than Trident-Hest), supporting scalable WSI preprocessing.

\subsubsection{Robustness across cohorts and artifacts}\label{sec:exp_robustness}

\begin{figure*}[pos=!htbp,width=\textwidth]
    \centering
    \includegraphics[width=0.7\textwidth]{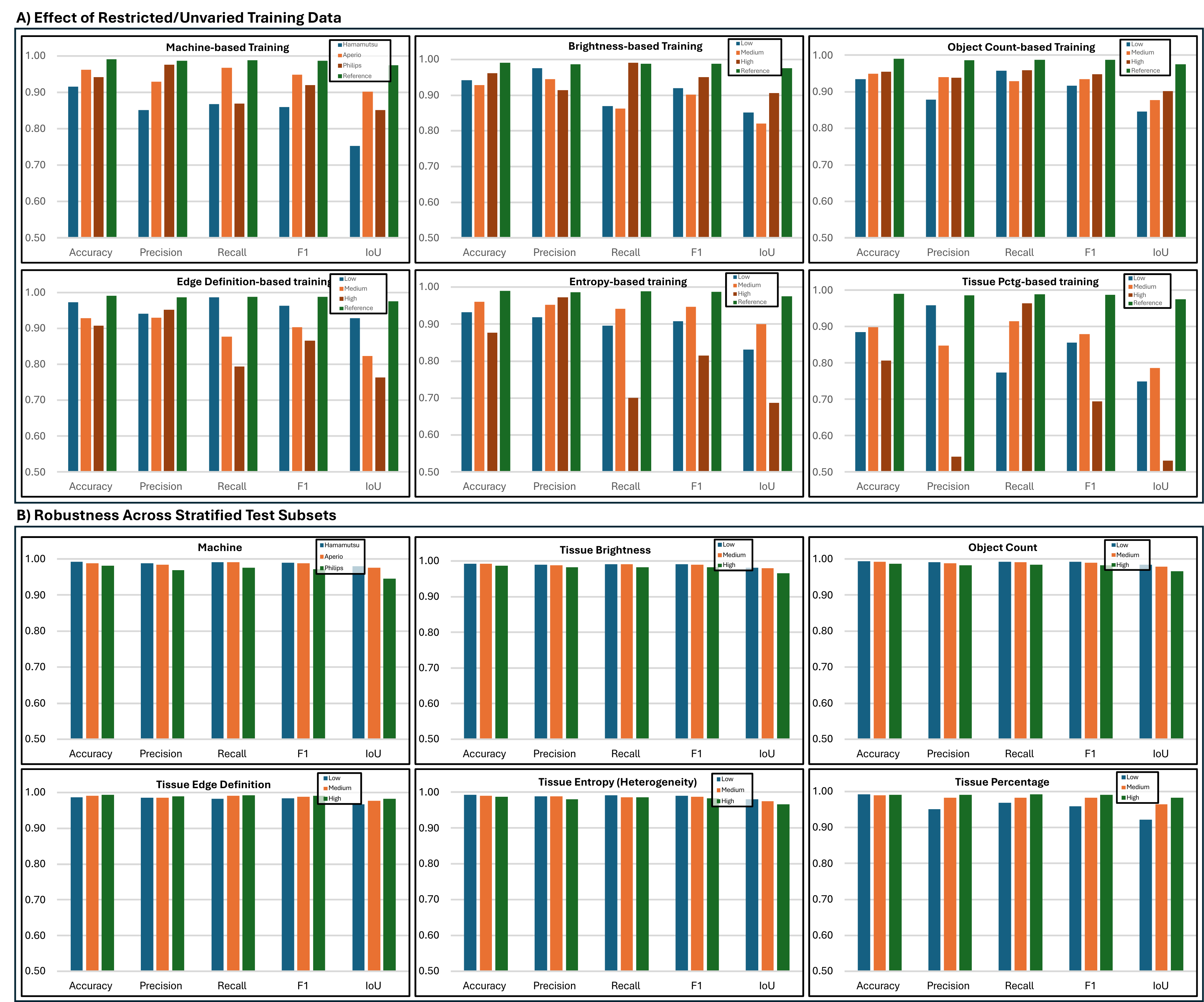}
    \caption{Multiple experiments to validate the importance of data diversity in the curated dataset. \textbf{A.}~Effect of restricted/unvaried training sets. Subsets of the original dataset are created for varying conditions (machines, tissue brightness, object count, etc). For each subset, the SAM2 model is finetuned on the given data, and tested on the remainder of the dataset. Results show training on a restricted set with no diversity cannot generalize to larger heterogeneous data, as shown by the six examples. \textbf{B.}~Model performance across varying test sets. The model finetuned on the full heterogeneous training set is tested on stratified subsets (unseen during training), each covering a certain level of a given feature. Results show more adaptability and robustness to variance.}
    \label{Fig5}
\end{figure*}

\begin{figure*}[pos=!htbp,width=\textwidth]
    \centering
    \includegraphics[width=0.7\textwidth]{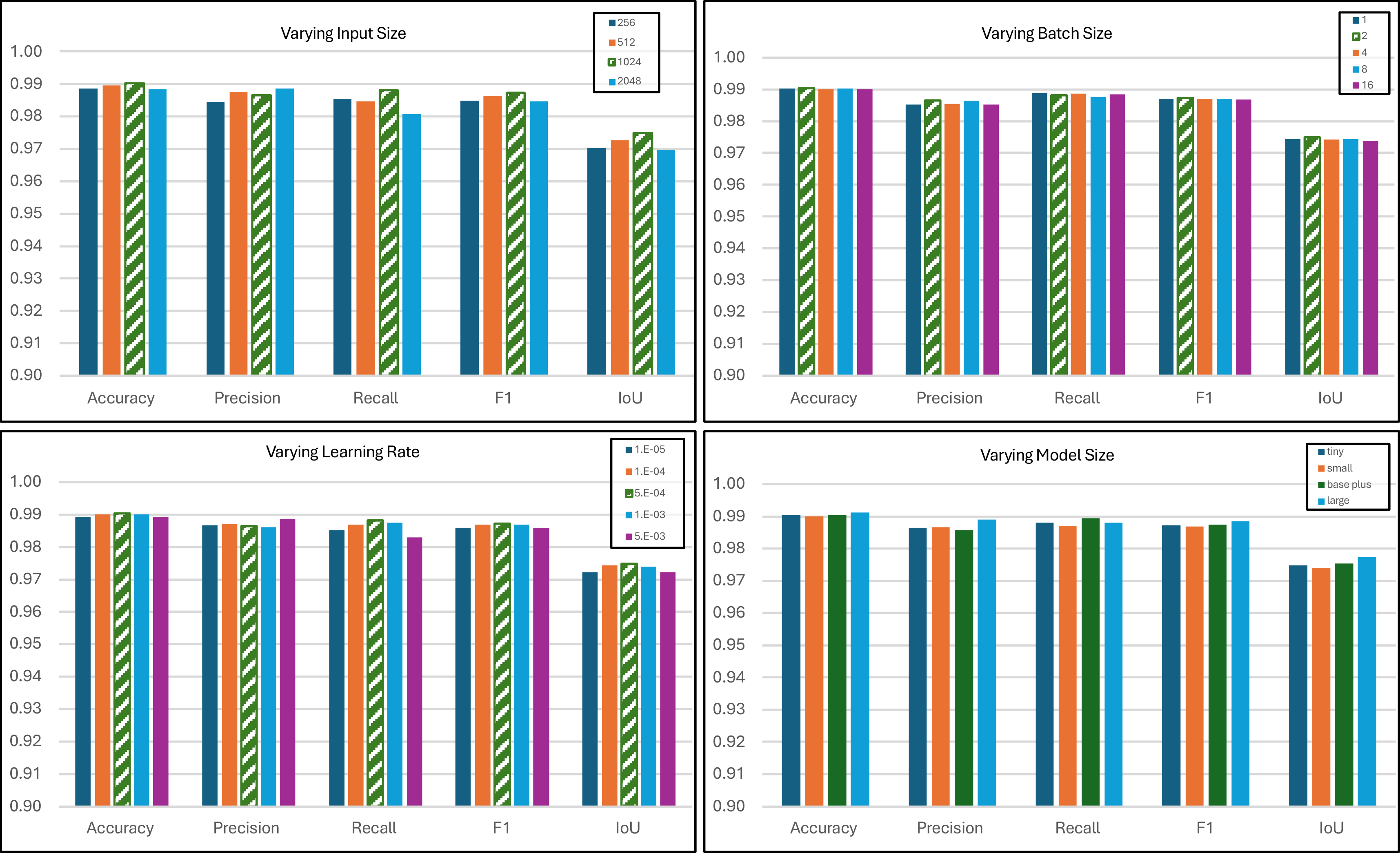}
    \caption{Ablative study on the tissue detection component in AtlasPatch under different training configurations and backbone size. Tissue detection performance is reported while varying key design choices, including input thumbnail resolution, training batch size, learning rate and the underlying SAM2 backbone variant. Across all tested settings, AtlasPatch maintains consistently high performance with only minor fluctuations, indicating that the method is robust to reasonable hyperparameter changes and that even lighter SAM2 backbones achieve competitive results, allowing practitioners to trade off memory and runtime against only marginal differences in segmentation quality.}
    \label{Fig6}
\end{figure*}

To test robustness, we created homogeneous training splits from our corpus by restricting data to a single scanner (Hamamatsu/Aperio/Philips) or to low/medium/high strata of key slide attributes (brightness, object count, edge definition, entropy and tissue percentage). We used each split to finetune SAM2 and evaluated the model on the remainder of the dataset outside that split (Fig.~\ref{Fig5}a). As expected, training on these narrow sets yields large performance swings across strata, with precision drops of up to 13.5\% for machine-, 7.3\% for brightness-, 10.8\% for object count-, 5.6\% for edge definition-, 6.7\% for entropy-, and 44.4\% for tissue percentage-based training. In contrast, when we train on the full heterogeneous multi-institutional corpus and then test on homogeneous subsets (unseen during training), the range (max--min) in precision is minimal (Fig.~\ref{Fig5}b), with 1.9\% for machine-, 0.71\% for brightness-, 0.86\% for object count-, 0.49\% for edge definition-, 0.88\% for entropy-, and 4.03\% for tissue percentage-based testing sets. The slightly larger range for tissue-percentage subsets is expected because in cases of slides with small tissues, even small extra predicted regions (false positives) make up a larger fraction of the predicted mask and therefore reduce precision more noticeably. These results validate our choice to train on a deliberately heterogeneous multi-cohort dataset, which is essential for reliable generalization across real-world WSI variability.

\subsection{Ablation studies}\label{sec:exp_ablation}

To assess sensitivity to implementation choices, we ablated AtlasPatch's training setup by varying input resolution (256--2048), batch size (1--16), learning rate ($1\times10^{-5}$ to $5\times10^{-3}$), and model size (tiny--large), with the default setup having input size of 1024, batch size of 2, learning rate of $5\times10^{-4}$, and tiny model size. The above variations produced only small changes in performance, with precision remaining between 0.984--0.989 and F1 between 0.985--0.988, indicating the tissue detector is robust to reasonable hyperparameter choices. The best overall trade-off is achieved at 1024$\times$1024, batch size 2, and learning rate $5\times10^{-4}$, which we adopted as our default. This is aligned with SAM2 native operating scale (1024 input size), avoiding unnecessary resampling. When varying the backbone from tiny (38.9M parameters) to small/base-plus/large (up to 224.4M), larger models yield marginal gains (precision and F1 increase by 0.002 and 0.001, respectively) while incurring $\sim$6$\times$ increase in parameter count and substantial runtime overhead, motivating SAM2.1\_hiera\_tiny as the default AtlasPatch backbone.

\subsection{Downstream MIL evaluation}\label{sec:exp_downstream}

We next assess the quality of the extracted patches on downstream slide-level prediction tasks via MIL while also reporting the computational complexity in terms of runtime, as shown in Fig.~\ref{Fig7}. To this end, we compare AtlasPatch to three representative slide-preprocessing pipelines: the two Trident variants (GrandQC and Hest) and CLAM. We separately use each tool to extract patch images for a given dataset to represent WSIs, and feed those with their labels to MIL to evaluate slide-level performance on six classification tasks. Performance is summarized using accuracy, precision, recall, F1 score, and ROC AUC, while monitoring the average number of extracted patches per slide for each tool. Across the six different downstream tasks, patches extracted via AtlasPatch achieve slide-level performance that is effectively on par with the best competing pipelines. Specifically, for the PANDA, LUADvsLUSC, and KIRCvsKIRP tasks, AtlasPatch achieves the highest performance with accuracies of 73.5\%, 95.8\%, and 97.7\%, respectively. For invasiveness and dysplasia, AtlasPatch achieves accuracies of 98.6\% and 96.6\%, respectively, remaining within at most 0.2\% of the highest performing pipeline. On BRCA, it also stays competitive, with 93.7\% accuracy compared to 94.4\% for CLAM, while using substantially fewer patches per WSI, nearly one third of those produced by CLAM containing a lot of background. Generally, we notice AtlasPatch achieves the aforementioned performance while producing, on average, lower number of patches per WSI ($\sim$3047 patches/WSI), compared to $\sim$8976, $\sim$3204, and $\sim$3092 patches/WSI for CLAM, GrandQC, and Hest, respectively, indicating it yields more focused and information-dense patches with less redundant background, thereby lowering the computational and storage burden on downstream models.

The plots in Fig.~\ref{Fig7} report the patch coordinate extraction runtime (inclusive of tissue detection) for 100 randomly sampled slides, and matches it with the MIL accuracy achieved in each task. When compared to the slide preprocessing tool in CLAM, AtlasPatch processes 100 slides for patch coordinate extraction in 195.51 seconds, which is more than 2$\times$ faster than CLAM and Trident-GrandQC, and more than 16$\times$ faster than Trident-Hest.

\FloatBarrier
\section{Discussion}\label{sec:discussion}

WSI preprocessing, being tissue detection followed by patch extraction, is often treated as a routine step that is taken for granted, yet it can become the dominant constraint when scaling to foundation-model training. AtlasPatch addresses this ``preprocessing tax'' by adopting a thumbnail-based design that detects tissue at low resolution and extrapolates contours to generate patch coordinates at the desired magnifications, avoiding repeated high-resolution reads and patch-level inference. Across qualitative stress cases, AtlasPatch produced consistent tissue masks under low contrast, heavy fragmentation, and common artifacts, conditions under which thresholding and patch-based deep learning pipelines often mishandle. AtlasPatch did not always reproduce every fine-grained internal hole within tissue regions, but it reliably recovered the full spatial footprint of diagnostically meaningful tissue and avoided large false negatives, as shown when we analyzed the model's precision. This efficiency is enabled by pairing the approach with a deliberately heterogeneous dataset and a parameter-efficient SAM2 adaptation strategy, which together improve robustness across varying conditions without incurring heavy training or deployment overhead. In contrast, pipelines that adopt deep segmenters at the patch level add complexity and compute through dense patch scanning and stitching, and their generalization can be constrained by limited training data (GrandQC \citep{weng2024grandqc} was trained on patches from only $\sim$600 WSIs) as illustrated in Fig.~\ref{Fig4}.

\begin{figure*}[pos=!htbp,width=\textwidth]
    \centering
    \includegraphics[width=\textwidth]{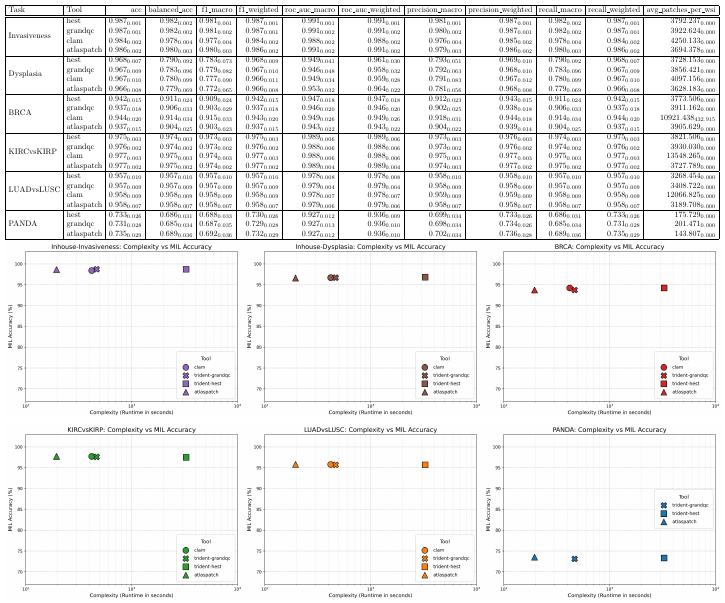}
    \caption{Downstream MIL performance and complexity of several end-to-end WSI preprocessing tools on 6 different tasks: colorectal cancer invasiveness (in-house), dysplasia level (in-house), breast carcinoma subtyping (TCGA-BRCA IDC vs ILC), renal cell carcinoma subtyping (TCGA-KIRC vs KIRP), lung adenocarcinoma versus squamous carcinoma (TCGA LUAD vs LUSC) and prostate cancer grading (PANDA). The table reports slide-level metrics and the average number of patch embeddings per WSI. For each task, we train several MIL methods on the patches extracted from each tool, and report the average results. The scatter plots show complexity--accuracy trade-offs for each task. Across all datasets, AtlasPatch achieves comparable or better downstream performance while using fewer embeddings on average and substantially lower runtime. Note: for the PANDA dataset, we exclude CLAM because it fails to detect tissue in $\sim$54\% of the slides.}
    \label{Fig7}
\end{figure*}

Importantly, AtlasPatch does not trade accuracy for speed where it matters most: downstream tasks. Across diverse MIL tasks (Fig.~\ref{Fig7}), AtlasPatch-derived patches achieve slide-level performance comparable to strong baselines while substantially reducing end-to-end runtime and redundant background-heavy sampling (more than twice as fast as Trident-GrandQC and 16$\times$ faster than Trident-Hest). In this sense, better tissue detection translates into a more focused ``search space'' for downstream models, which is especially relevant when storage, runtime, and encoder time become limiting factors at scale. Beyond performance, AtlasPatch was designed for ease of use by both researchers and pathologists. The pipeline is presented as four modules, each with its checkpoint and human-readable commands (e.g., tissue visualization only, coordinate extraction, full processing with patch embeddings, and optional patch export), allowing users to ``check out'' at any stage depending on whether they need quick QC overlays, patch coordinates for MIL, or ready-to-train embeddings. Outputs are standardized (mask/contour/grid visualizations for rapid review, and per-slide HDF5 files containing coordinates and features), enabling straightforward integration into existing analysis workflows and reproducible reruns. Practical deployment is supported through simple single-slide or directory-level execution, explicit CPU/GPU controls, and robust run management (e.g., skipping completed slides and resumable outputs), alongside optimized parallel execution across modules for patch I/O, model calls, and coordinate extraction on CPU and GPU when applicable.

AtlasPatch's efficiency and modularity are especially valuable as computational pathology pivots toward foundation models, which require unprecedented data throughput and flexible preprocessing at scale \citep{xiong2025survey}. Major foundation models exemplify this trend with massive dataset sizes, such as HIPT \citep{Chen2022Scaling} ($\sim$104M image patches), RudolfV \citep{dippel2024rudolfv} ($\sim$134k WSIs $\approx$ 1.2B patches), PLIP \citep{huang2023visual} ($\sim$208k image-caption pairs), UNI \citep{chen2024towards} ($\sim$100k WSIs $\approx$ 100M patches), and CONCH \citep{lu2024visual} (more than 1.17M image-caption pairs). This scale comes with formidable runtime costs, as repeatedly extracting patches at multiple magnifications or patch sizes becomes a severe I/O and computation bottleneck at such magnitudes. This is only the tip of the iceberg, as upcoming years will likely demand even larger models and bigger datasets, further amplifying the strain on preprocessing pipelines. Efficient and scalable tools like AtlasPatch will be essential to keep pace with these demands and unlock the full potential of foundation models in pathology.

Although AtlasPatch streamlines WSI tissue segmentation and patch extraction, manual annotation remains a key bottleneck that required substantial human effort and manpower. One practical future direction is to incorporate active learning, which helps prioritize uncertain or diverse WSI thumbnails for labeling to reduce the amount of annotation needed \citep{meirelles2022effective}. This opens the door to even larger and more diverse cohorts for training tissue detectors. Another potential direction is the tool integration into routine clinical QC workflows. While AtlasPatch is already designed to be easy to use, with modular stages, clear checkpoints, and standardized visual overlays, wider adoption will benefit from dedicated pathologist-friendly GUI components and tighter integration with slide viewers and QC tools, enabling rapid review, flagging, and lightweight corrections that can be fed back into model updates.

\section{Conclusion}\label{sec:conclusion}

We presented AtlasPatch, a scalable WSI preprocessing framework that couples thumbnail-based tissue detection with high-throughput patch extraction and embedding. By adapting SAM2 through parameter-efficient fine-tuning on a large, heterogeneous multi-cohort dataset of approximately 30,000 annotated WSI thumbnails, AtlasPatch achieves a tissue-detection precision of 0.986 while remaining robust across variations in brightness, fragmentation, tissue heterogeneity, boundary definition, and common artifacts such as pen markings and scanner streaks. Compared with widely used preprocessing pipelines, AtlasPatch reduces end-to-end preprocessing time by up to 16$\times$ without degrading downstream MIL performance across six slide-level classification tasks spanning multiple organs and cohorts. The modular, parallelized design, supporting independent tissue detection, coordinate extraction, embedding, and optional patch export, makes the tool adaptable to both clinical quality-control workflows and large-scale AI research pipelines. As computational pathology increasingly demands foundation-model-scale data throughput, efficient and robust preprocessing tools like AtlasPatch will be essential to keep pace with these growing requirements.

\section{Data and Code Availability}\label{sec:availability}

TCGA data can be downloaded from the GDC platform (\url{https://portal.gdc.cancer.gov/}). The CAMELYON17 dataset is available on the Grand Challenge data page (\url{https://camelyon17.grand-challenge.org/Data/}). The PANDA dataset can be downloaded from the Kaggle competition page (\url{https://www.kaggle.com/c/prostate-cancer-grade-assessment/data}). In-house CHUM cohorts (pancreas and digestive-system) were used under institutional approvals and data-sharing agreements and are not publicly available.

AtlasPatch is open-sourced at \url{https://github.com/AtlasAnalyticsLab/AtlasPatch}. The trained tissue detection model checkpoint is hosted on Hugging Face at \url{https://huggingface.co/AtlasAnalyticsLab/AtlasPatch}. AtlasPatch automatically downloads/loads these weights from Hugging Face, but users must provide a valid Hugging Face access token to enable model retrieval. The use for commercial purposes is not permitted.

AtlasPatch is distributed as a Python package with services for slide loading, tissue segmentation, contour-based coordinate extraction, visualization, storage, and HDF5 export. It supports OpenSlide-backed pathology formats such as SVS, TIFF, NDPI, MRXS, BIF, and DICOM, as well as standard images. The repository also provides per-slide microns-per-pixel overrides, patch ``passport'' identifiers, PNG patch export, quality-control overlays, lock-based concurrency control, and SLURM templates for staged preprocessing on compute clusters. 

The encoder registry covers natural-image backbones, pathology foundation models, and medical multimodal encoders, reaching 65+ patch encoder. AtlasPatch produces embeddings from 384 to 4096 dimensions and stores them in the HDF5 \texttt{features} group aligned with \texttt{coords}. Multiple encoders can be requested in one run, with control over device placement, batch size, worker count, and numerical precision (\texttt{float32}, \texttt{float16}, or \texttt{bfloat16}). Custom patch encoders can be added and used through the same command interface without modifying the core package.

\appendix
\setcounter{figure}{0}
\renewcommand{\thefigure}{A.\arabic{figure}}
\renewcommand{\theHfigure}{appendix.\arabic{figure}}

\section*{Appendix A: Supplementary Figures}\label{sec:appendix}

\begin{figure*}[pos=!htbp,width=\textwidth]
    \centering
    \includegraphics[width=\textwidth]{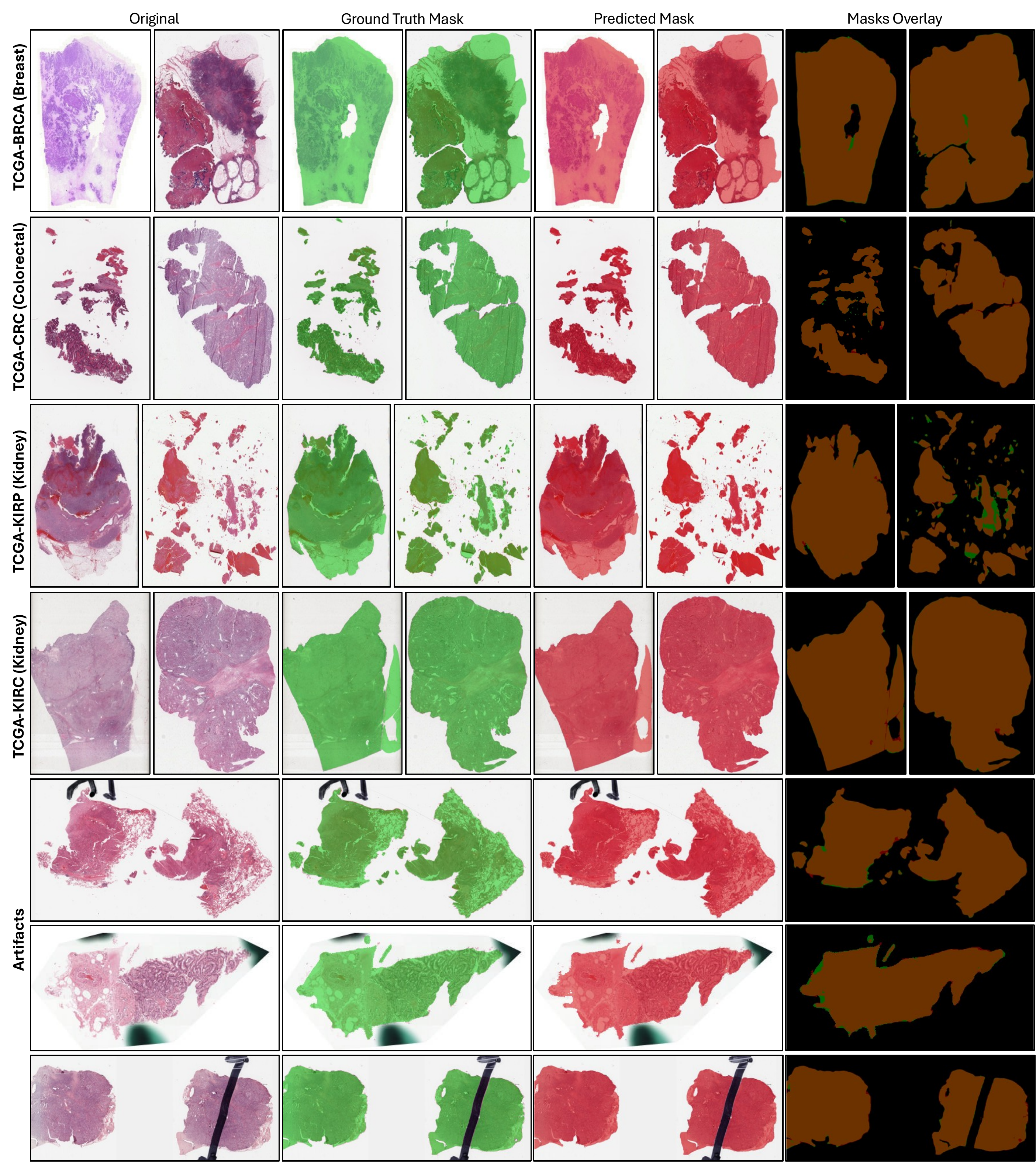}
    \caption{More qualitative examples to support Fig.~\ref{Fig3}, showing original slide thumbnails along with the ground truth annotation mask (overlayed on the slide in green), the model's predicted mask (overlayed in red), and an overlay of both masks to show intersection.}
    \label{Fig3supp}
\end{figure*}

\begin{figure*}[pos=!htbp,width=\textwidth]
    \centering
    \includegraphics[width=\textwidth]{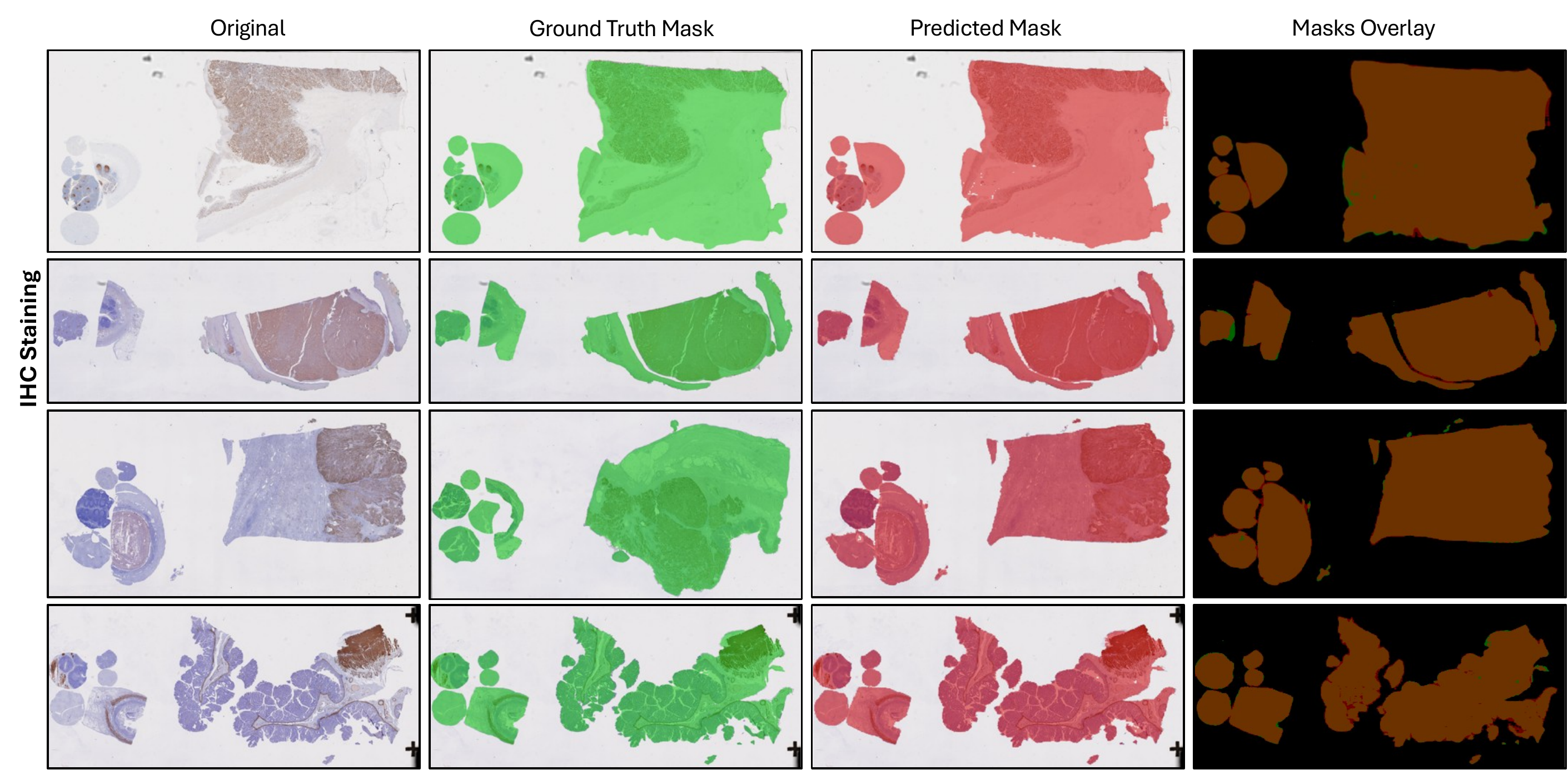}
    \caption{More qualitative examples to support Fig.~\ref{Fig3}, specifically for samples with IHC staining, showing the generalization capability of the tissue detector in AtlasPatch.}
    \label{Fig3supp2}
\end{figure*}

\begin{figure*}[pos=!htbp,width=\textwidth]
    \centering
    \includegraphics[width=\textwidth]{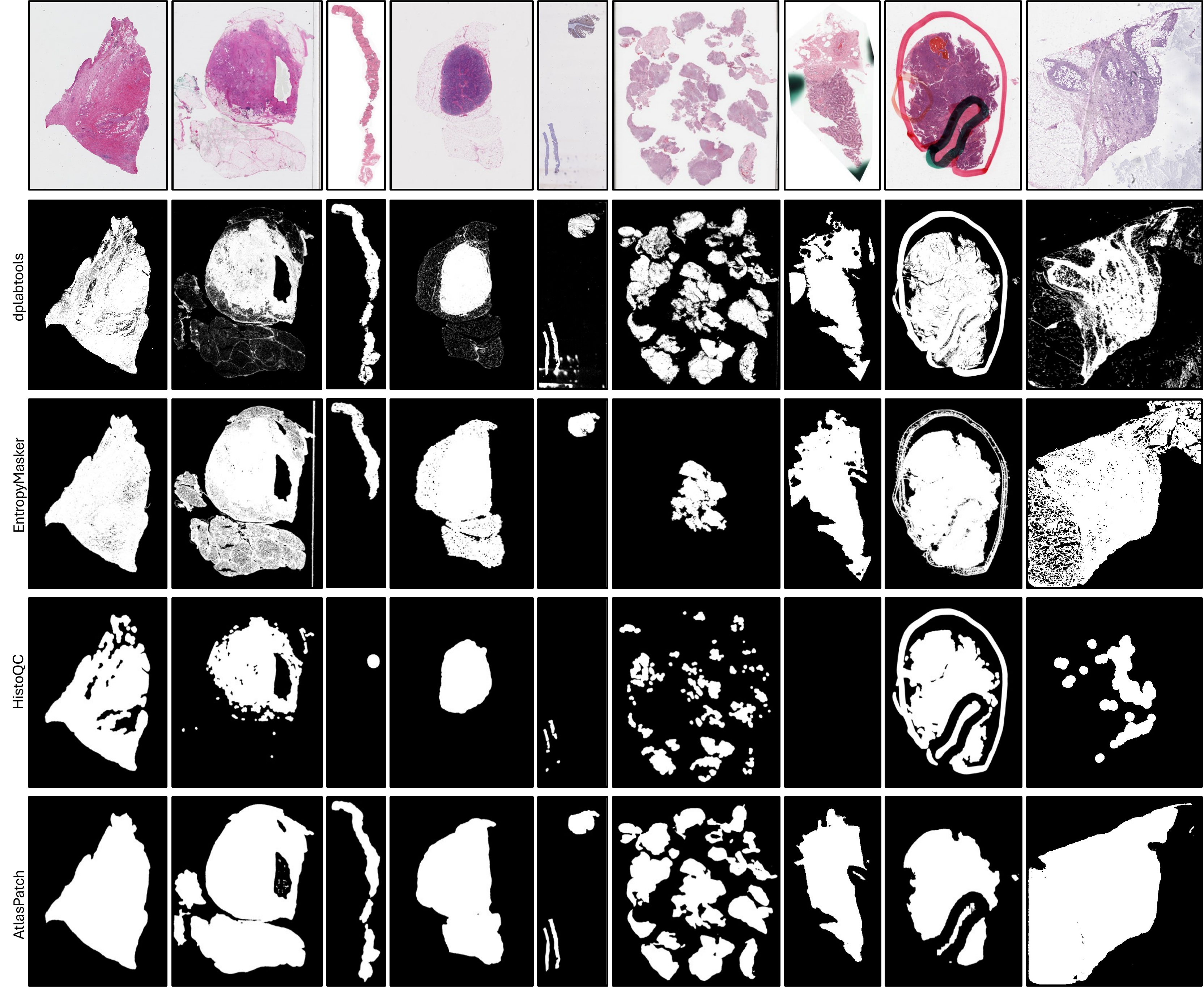}
    \caption{More qualitative examples to support Fig.~\ref{Fig4}, showing original slide thumbnails along with the predicted masks of three thresholding tools, in addition to AtlasPatch.}
    \label{Fig4supp}
\end{figure*}

\begin{figure*}[pos=!htbp,width=\textwidth]
    \centering
    \includegraphics[width=\textwidth]{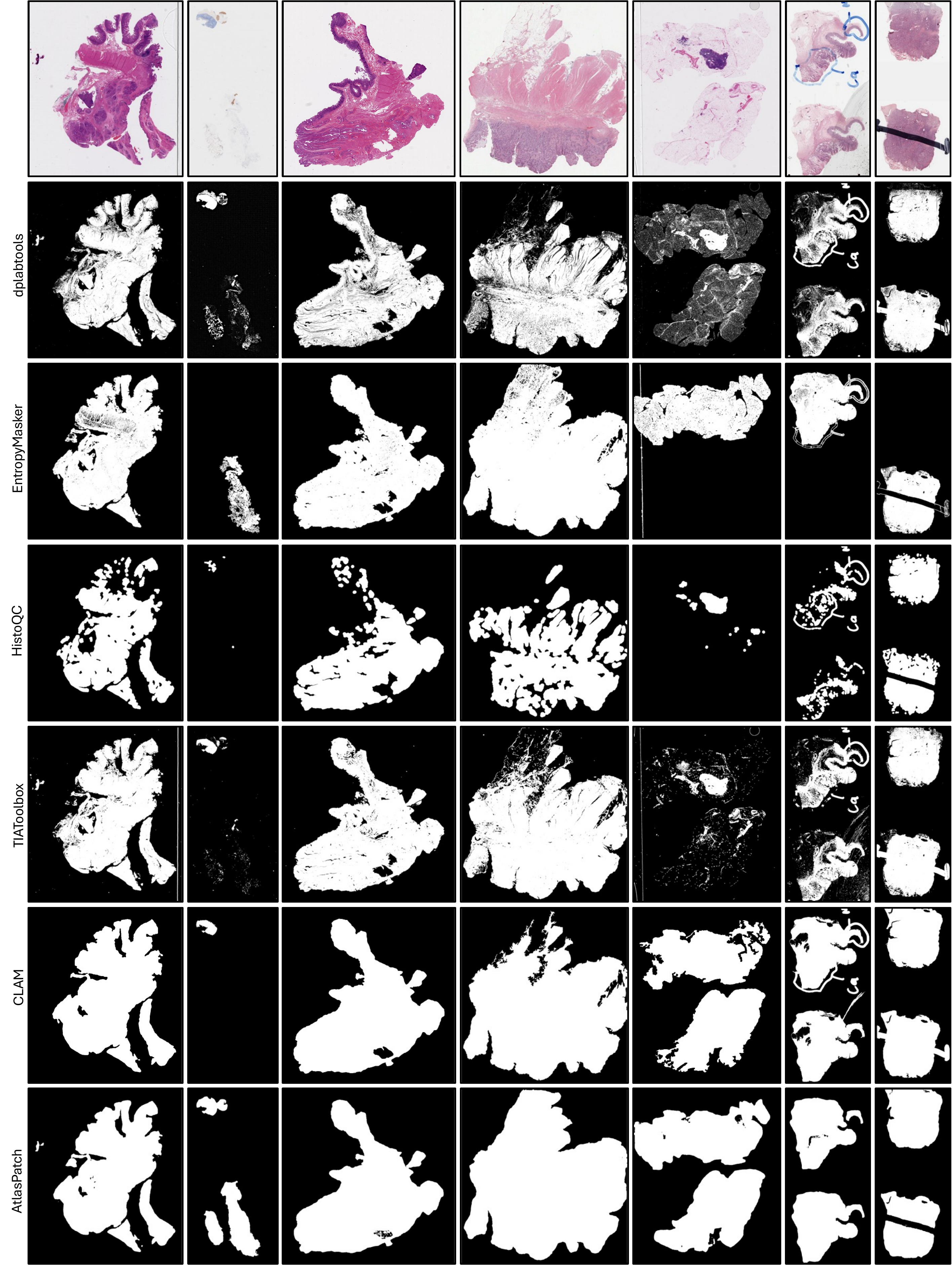}
    \caption{More qualitative examples to support Fig.~\ref{Fig4} for thresholding-based methods in comparison with AtlasPatch.}
    \label{Fig4supp2}
\end{figure*}

\begin{figure*}[pos=!htbp,width=\textwidth]
    \centering
    \includegraphics[width=\textwidth]{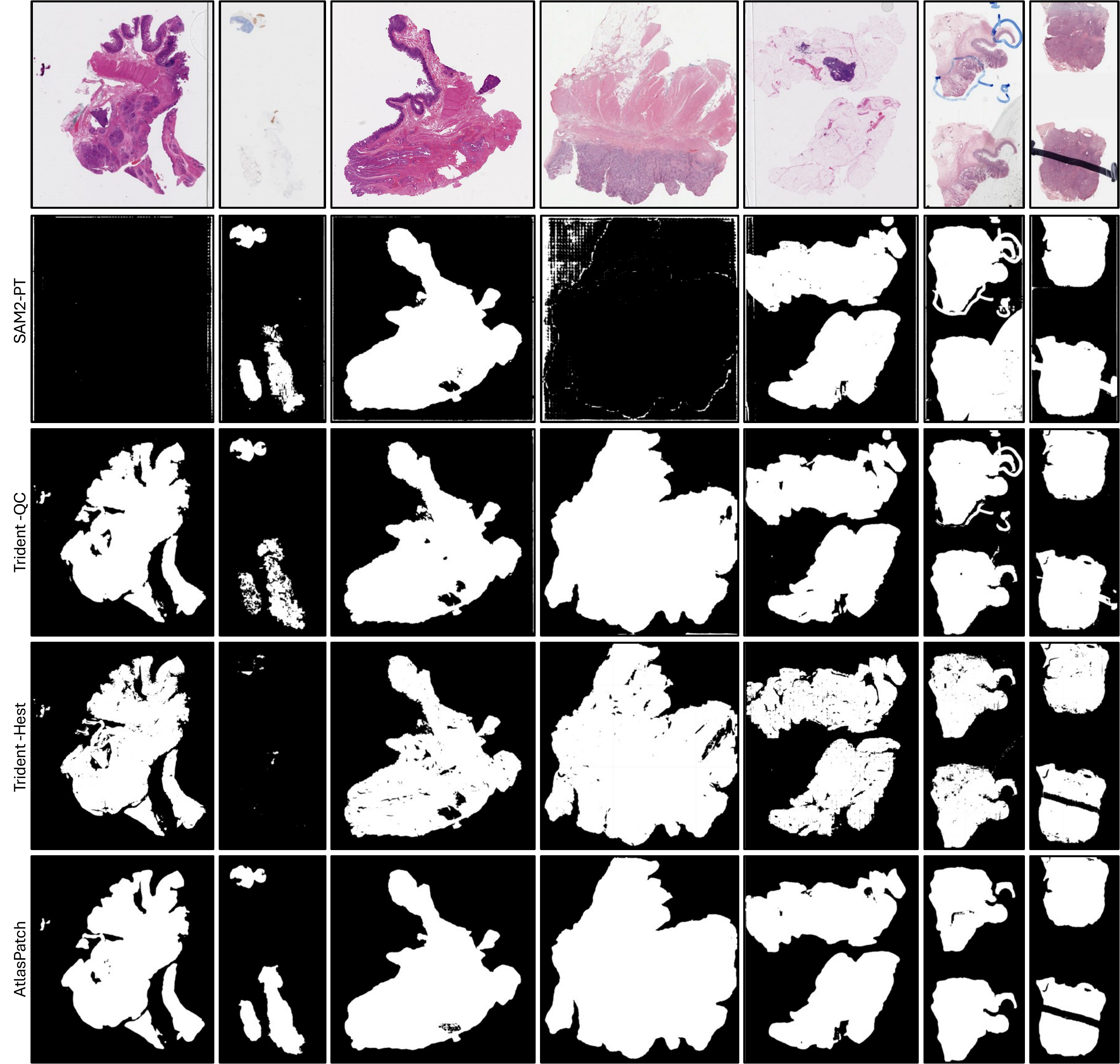}
    \caption{More qualitative examples to support Fig.~\ref{Fig4} for AI-based methods in comparison with AtlasPatch.}
    \label{Fig4supp3}
\end{figure*}

This appendix provides additional qualitative results that complement the main-text evaluations. Figures~\ref{Fig3supp} and~\ref{Fig3supp2} extend Fig.~\ref{Fig3} with more examples of AtlasPatch tissue-detection performance, including IHC-stained cases, while Figs.~\ref{Fig4supp}--\ref{Fig4supp3} extend Fig.~\ref{Fig4} with further comparisons against thresholding-based and AI-based preprocessing baselines.

\FloatBarrier
\clearpage

\printcredits

\section*{Acknowledgments}
This research was partially supported by the following grants: NSERC-DG RGPIN-2022-05378 [M.S.H], Amazon Research Award [M.S.H], Gina Cody RIF [M.S.H], the Canadian Cancer Society Breakthrough Grant [V.Q.H.T], FRQS-CRS-J1 [V.Q.H.T], Institute for Research in Immunology and Cancer start up funds [V.Q.H.T], and FRQNT scholarships [A.A and Y.K]. The experiments were enabled in part by support provided by Calcul Quebec (\url{www.calculquebec.ca}) and the Digital Research Alliance of Canada (\url{www.alliance.can.ca}).

\bibliographystyle{cas-model2-names}
\bibliography{cas-refs}

@article{bandi2018detection,
  title={From detection of individual metastases to classification of lymph node status at the patient level: the camelyon17 challenge},
  author={Bandi, Peter and Geessink, Oscar and Manson, Quirine and Van Dijk, Marcory and Balkenhol, Maschenka and Hermsen, Meyke and Babak Ehteshami Bejnordi and Byungjae Lee and Kyunghyun Paeng and Aoxiao Zhong and Quanzheng Li and Farhad Ghazvinian Zanjani and Svitlana Zinger and Keisuke Fukuta and Daisuke Komura and Vlado Ovtcharov and Shenghua Cheng and Shaoqun Zeng and Jeppe Thagaard and Anders B. Dahl and Huangjing Lin and Hao Chen and Ludwig Jacobsson and Martin Hedlund and Melih Cetin and Eren Halıcı and Hunter Jackson and Richard Chen and Fabian Both and Jorg Franke and Heidi Kusters-Vandevelde and Willem Vreuls and Peter Bult and Bram van Ginneken and Jeroen van der Laak and Geert Litjens},
  journal={IEEE transactions on medical imaging},
  volume={38},
  number={2},
  pages={550--560},
  year={2018},
  publisher={IEEE}
}

@article{bulten2022artificial,
  title={Artificial intelligence for diagnosis and Gleason grading of prostate cancer: the PANDA challenge},
  author={Bulten, Wouter and Kartasalo, Kimmo and Chen, Po-Hsuan Cameron and Str{\"o}m, Peter and Pinckaers, Hans and Nagpal, Kunal and Cai, Yuannan and Steiner, David F and Van Boven, Hester and Vink, Robert and Christina Hulsbergen-van de Kaa and Jeroen van der Laak and Mahul B. Amin and Andrew J. Evans and Theodorus van der Kwast and Robert Allan and Peter A. Humphrey and Henrik Grönberg and Hemamali Samaratunga and Brett Delahunt and Toyonori Tsuzuki and Tomi Häkkinen and Lars Egevad and Maggie Demkin and {the PANDA challenge consortium}},
  journal={Nature medicine},
  volume={28},
  number={1},
  pages={154--163},
  year={2022},
  publisher={Nature Publishing Group US New York}
}

@article{weinstein2013cancer,
  title={The cancer genome atlas pan-cancer analysis project},
  author={Weinstein, John N and Collisson, Eric A and Mills, Gordon B and Shaw, Kenna R and Ozenberger, Brad A and Ellrott, Kyle and Shmulevich, Ilya and Sander, Chris and Stuart, Joshua M},
  journal={Nature genetics},
  volume={45},
  number={10},
  pages={1113--1120},
  year={2013},
  publisher={Nature Publishing Group}
}

@article{Janowczyk2019HistoQC,
  author  = {Andrew Janowczyk and Ren Zuo and Hannah Gilmore and Michael D. Feldman and Anant Madabhushi},
  title   = {{HistoQC}: An Open-Source Quality Control Tool for Digital Pathology Slides},
  journal = {JCO Clinical Cancer Informatics},
  year    = {2019},
  volume  = {3},
  pages   = {1--7},
  doi     = {10.1200/CCI.18.00157},
  url     = {https://ascopubs.org/doi/10.1200/CCI.18.00157}
}

@misc{HistoQCGit,
  author  = {Janowczyk, Andrew and contributors},
  title   = {HistoQC},
  howpublished = {\url{https://github.com/choosehappy/HistoQC}},
}

@misc{EntropyMaskerGit,
  author  = {Song, Yipei and contributors},
  title   = {EntropyMasker},
  howpublished = {\url{https://github.com/CirculatoryHealth/EntropyMasker}},
}

@article{Berman2021PathML,
  author  = {Jacob Rosenthal and Ryan Carelli and Mohamed Omar and David Brundage and Ella Halbert and Jackson Nyman and Surya N. Hari and Eliezer M. Van Allen and Luigi Marchionni and Renato Umeton and Massimo Loda},
  title   = {Building Tools for Machine Learning and Artificial Intelligence in Cancer Research: Best Practices and a Case Study with the {PathML} Toolkit for Computational Pathology},
  journal = {Molecular Cancer Research},
  year    = {2022},
  volume  = {20},
  number  = {2},
  pages   = {202--206},
  doi     = {10.1158/1541-7786.MCR-21-0665},
  url     = {https://aacrjournals.org/mcr/article/20/2/202/678062/}
}

@misc{PathMLDocs,
  author = {{PathML Developers}},
  title = {{PathML} Documentation},
  year = {2026},
  url = {https://pathml.readthedocs.io/},
  note = {Documentation website, accessed 2026-03-23}
}

@article{PathEX2024,
  title   = {PathEX: Make Good Choice for Whole Slide Image Extraction},
  author = {Yang X. and Zhang R. and Yang Y. and Zhang Y. and Chen K.},
  journal = {PLOS ONE},
  year    = {2024},
  volume  = {19},
  number  = {8},
  pages   = {e0304702},
  doi     = {10.1371/journal.pone.0304702},
  url     = {https://journals.plos.org/plosone/article?id=10.1371/journal.pone.0304702}
}

@misc{Lu2021CLAM,
  author        = {Ming Y. Lu and Drew F. K. Williamson and Tiffany Y. Chen and Richard J. Chen and Matteo Barbieri and Faisal Mahmood},
  title         = {Data-efficient and weakly supervised computational pathology on whole-slide images},
  year          = {2020},
  eprint        = {2004.09666},
  archivePrefix = {arXiv},
  primaryClass  = {eess.IV},
  url           = {https://arxiv.org/abs/2004.09666}
}

@misc{CLAMGit,
  author = {{Mahmood Lab}},
  title = {{CLAM}: Open-source tools for computational pathology on {WSIs}},
  year = {2021},
  url = {https://github.com/mahmoodlab/CLAM},
  note = {GitHub repository, accessed 2025-08-23}
}

@misc{TRIDENTDocs,
  author = {{Mahmood Lab}},
  title = {{TRIDENT} Documentation},
  year = {2025},
  url = {https://trident-docs.readthedocs.io/},
  note = {Documentation website, accessed 2025-08-28}
}

@misc{TRIDENTGit,
  author = {{Mahmood Lab}},
  title = {{TRIDENT}: Toolkit for large-scale whole-slide image processing},
  year = {2025},
  url = {https://github.com/mahmoodlab/TRIDENT},
  note = {GitHub repository, accessed 2025-08-23}
}

@misc{TIADocs,
  author = {{Tissue Image Analytics Centre}},
  title = {{TIAToolbox} Documentation},
  year = {2025},
  url = {https://tia-toolbox.readthedocs.io/},
  note = {Documentation website, accessed 2025-08-16}
}

@misc{OpenSlideSite,
  author = {{OpenSlide Team}},
  title = {{OpenSlide}},
  year = {2026},
  url = {https://openslide.org/},
  note = {Project website, accessed 2026-03-23}
}

@misc{OpenSlideGit,
  author = {{OpenSlide Team}},
  title = {{OpenSlide-Python}},
  year = {2026},
  url = {https://github.com/openslide/openslide-python},
  note = {GitHub repository, accessed 2026-03-23}
}

@article{goode2013openslide,
  title={OpenSlide: A vendor-neutral software foundation for digital pathology},
  author={Goode, Adam and Gilbert, Benjamin and Harkes, Jan and Jukic, Drazen and Satyanarayanan, Mahadev},
  journal={Journal of pathology informatics},
  volume={4},
  number={1},
  pages={27},
  year={2013},
  publisher={Elsevier}
}

@article{DPLabPaper,
  author  = {Alice Shen and Fusheng Wang and Saptarshi Paul and Divya Bhuvanapalli and Jacob Alayof and Alton B. Farris and George Teodoro and Daniel J. Brat and Jun Kong},
  title   = {An integrative web-based software tool for multi-dimensional pathology whole-slide image analytics},
  journal = {Physics in Medicine \& Biology},
  year    = {2022},
  volume  = {67},
  number  = {22},
  pages   = {224001},
  doi     = {10.1088/1361-6560/ac8fde},
  url     = {https://pubmed.ncbi.nlm.nih.gov/36067783/}
}

@misc{DPLabDocs,
  author = {{Sunnybrook Research Institute}},
  title = {Digital Pathology Lab Tools Documentation},
  year = {2024},
  url = {https://dplabtools.readthedocs.io/en/latest/},
  note = {Version 0.0.4 documentation, accessed 2025-09-23}
}

@misc{SlideflowGit,
  author = {{Slideflow Developers}},
  title = {{Slideflow}},
  year = {2024},
  url = {https://github.com/slideflow/slideflow},
  note = {GitHub repository, accessed 2025-09-23}
}

@inproceedings{Ronneberger2015UNet,
  author    = {Olaf Ronneberger and Philipp Fischer and Thomas Brox},
  title     = {{U-Net: Convolutional Networks for Biomedical Image Segmentation}},
  booktitle = {Medical Image Computing and Computer-Assisted Intervention (MICCAI)},
  series    = {LNCS},
  volume    = {9351},
  pages     = {234--241},
  year      = {2015},
  organization = {Springer}
}

@article{Alsaafin2024SPLICE,
  author    = {Areej Alsaafin and Peyman Nejat and Abubakr Shafique and Jibran Khan and Saghir Alfasly and Ghazal Alabtah and Hamid R. Tizhoosh},
  title     = {{Sequential Patching Lattice for Image Classification and Enquiry: Streamlining Digital Pathology Image Processing}},
  journal   = {American Journal of Pathology},
  volume    = {194},
  number    = {10},
  pages     = {1898--1912},
  year      = {2024},
  doi       = {10.1016/j.ajpath.2024.06.007}
}

@article{Zhang2025Trident,
  author    = {Andrew Zhang and Guillaume Jaume and Anurag Vaidya and Tong Ding and Faisal Mahmood},
  title     = {{Accelerating Data Processing and Benchmarking of AI Models for Pathology}},
  journal   = {arXiv preprint arXiv:2502.06750},
  year      = {2025},
  note      = {https://github.com/mahmoodlab/trident}
}

@article{Bankhead2017QuPath,
  author    = {Peter Bankhead and Maurice B. Loughrey and Jos\'{e} A. Fern\'{a}ndez and Yvonne Dombrowski and Darragh G. Dunne and Stephen McArt and Ronan T. Gray and Liam J. Murray and Heather G. Coleman and Jim A. James and Manuel Salto-Tellez and Peter W. Hamilton},
  title     = {{QuPath: Open source software for digital pathology image analysis}},
  journal   = {Scientific Reports},
  volume    = {7},
  pages     = {16878},
  year      = {2017},
  doi       = {10.1038/s41598-017-17204-5}
}

@article{pocock2022tiatoolbox,
  title={TIAToolbox as an end-to-end library for advanced tissue image analytics},
  author={Pocock, Johnathan and Graham, Simon and Vu, Quoc-Dang and Jahanifar, Mostafa and Deshpande, Srijay and Hadjigeorghiou, Giorgos and Shephard, Adam and Smith, Kevin and Raza, Shan E Ahmed and Minhas, Fayyaz ul Amir Afsar and Rajpoot, Nasir},
  journal={Medical Image Analysis},
  volume={76},
  pages={102305},
  year={2022},
  publisher={Elsevier}
}

@article{dolezal2024slideflow,
  title={Slideflow: deep learning for digital histopathology with real-time whole-slide visualization},
  author={Dolezal, James M and Kochanny, Sara and Dyer, Emma and Ramesh, Siddhi and Srisuwananukorn, Andrew and Sacco, Matteo and Howard, Frederick M and Li, Anran and Mohan, Prajval and Pearson, Alexander T},
  journal={BMC bioinformatics},
  volume={25},
  number={1},
  pages={134},
  year={2024},
  publisher={Springer}
}

@article{ciga2021patchlimit,
  title={Overcoming the limitations of patch-based learning to detect cancer in whole-slide images},
  author={Ciga, Ozan and Xu, Tony and Martel, Anne L},
  journal={Scientific Reports},
  volume={11},
  number={1},
  pages={4557},
  year={2021},
  publisher={Nature Publishing Group}
}

@inproceedings{otsu1979threshold,
  title={A threshold selection method from gray-level histograms},
  author={Otsu, Nobuyuki},
  booktitle={IEEE Transactions on Systems, Man, and Cybernetics},
  volume={9},
  number={1},
  pages={62--66},
  year={1979},
  organization={IEEE}
}

@article{Hanna2022Integrating,
  author    = {Matthew G. Hanna and Orly Ardon and Victor E. Reuter and Sahussapont J. Sirintrapun and Christine England and David S. Klimstra and Meera R. Hameed},
  title     = {Integrating digital pathology into clinical practice},
  journal   = {Modern Pathology},
  volume    = {35},
  number    = {2},
  pages     = {152--164},
  year      = {2022},
  doi       = {10.1038/s41379-021-00929-0}
}

@article{Zia2025Update,
  author    = {Zia, Shamail and Yildiz-Aktas, Isil Z and Zia, Fazail and Parwani, Anil V},
  title     = {An update on applications of digital pathology: primary diagnosis; telepathology, education and research},
  journal   = {Diagnostic Pathology},
  volume    = {20},
  pages     = {17},
  year      = {2025},
  doi       = {10.1186/s13000-025-01610-9}
}

@article{hosseini2024computational,
  title={Computational pathology: a survey review and the way forward},
  author={Hosseini, Mahdi S and Bejnordi, Babak Ehteshami and Trinh, Vincent Quoc-Huy and Chan, Lyndon and Hasan, Danial and Li, Xingwen and Yang, Stephen and Kim, Taehyo and Zhang, Haochen and Wu, Theodore and Kajanan Chinniah and Sina Maghsoudlou and Ryan Zhang and Jiadai Zhu and Samir Khaki and Andrei Buin and Fatemeh Chaji and Ala Salehi and Bich Ngoc Nguyen and Dimitris Samaras and Konstantinos N. Plataniotis},
  journal={Journal of Pathology Informatics},
  volume={15},
  pages={100357},
  year={2024},
  publisher={Elsevier}
}

@article{weng2024grandqc,
  title={GrandQC: A comprehensive solution to quality control problem in digital pathology},
  author={Weng, Zhilong and Seper, Alexander and Pryalukhin, Alexey and Mairinger, Fabian and Wickenhauser, Claudia and Bauer, Marcus and Glamann, Lennert and Bl{\"a}ker, Hendrik and Lingscheidt, Thomas and Hulla, Wolfgang and Danny Jonigk and Simon Schallenberg and Andrey Bychkov and Junya Fukuoka and Martin Braun and Birgid Schomig-Markiefka and Sebastian Klein and Andreas Thiel and Katarzyna Bozek and George J. Netto and Alexander Quaas and Reinhard Buttner and Yuri Tolkach},
  journal={Nature Communications},
  volume={15},
  number={1},
  pages={10685},
  year={2024},
  publisher={Nature Publishing Group UK London}
}

@inproceedings{gu2018multi,
  title={Multi-resolution networks for semantic segmentation in whole slide images},
  author={Gu, Feng and Burlutskiy, Nikolay and Andersson, Mats and Wil{\'e}n, Lena Kajland},
  booktitle={International Workshop on Ophthalmic Medical Image Analysis},
  pages={11--18},
  year={2018},
  organization={Springer}
}

@inproceedings{Sommer2011,
  author    = {Sommer, Christoph and Straehle, Christoph and K{\"o}the, Ullrich and Hamprecht, Fred A.},
  title     = {ilastik: Interactive learning and segmentation toolkit},
  booktitle = {ISBI},
  pages     = {230--233},
  year      = {2011},
  doi       = {10.1109/ISBI.2011.5872394}
}

@inproceedings{Bandi2017Comparison,
  author    = {Bandi, Peter and van de Loo, Rob and Intezar, Milad and Geijs, Dion and Ciompi, Francesco and van Ginneken, Bram and van der Laak, Jeroen and Litjens, Geert},
  title     = {Comparison of different methods for tissue segmentation in histopathological whole-slide images},
  booktitle = {ISBI},
  pages     = {591--595},
  year      = {2017}
}

@article{Bandi2019,
  author  = {Bandi, Peter and Balkenhol, Michel and van Ginneken, Bram and van der Laak, Jeroen and Litjens, Geert},
  title   = {Resolution-agnostic tissue segmentation in whole-slide histopathology images with convolutional neural networks},
  journal = {PeerJ},
  volume  = {7},
  pages   = {e8242},
  year    = {2019}
}

@inproceedings{Kirillov2023SAM,
  title={Segment anything},
  author={Kirillov, Alexander and Mintun, Eric and Ravi, Nikhila and Mao, Hanzi and Rolland, Chloe and Gustafson, Laura and Xiao, Tete and Whitehead, Spencer and Berg, Alexander C and Lo, Wan-Yen and Piotr Dollar and Ross Girshick},
  booktitle={Proceedings of the IEEE/CVF international conference on computer vision},
  pages={4015--4026},
  year={2023}
}

@article{Zhang2023SAMPath,
  author  = {Zhang, Jingwei and Ma, Ke and Kapse, Saarthak and Saltz, Joel and Blais, Michael and Wang, Fusheng},
  title   = {{SAM-Path}: A Segment Anything Model for Semantic Segmentation in Digital Pathology},
  journal = {arXiv preprint arXiv:2307.09570},
  year    = {2023}
}

@inproceedings{Chen2022Scaling,
  author    = {Chen, Richard J. and Chen, Chengkuan and Li, Yicong and Chen, Tiffany Y. and Trister, Andrew and Krishnan, Rahul G. and Mahmood, Faisal},
  title     = {Scaling Vision Transformers to Gigapixel Images via Hierarchical Self-Supervised Learning},
  booktitle = {CVPR},
  year      = {2022}
}

@inproceedings{zhou2018unet++,
  title={Unet++: A nested u-net architecture for medical image segmentation},
  author={Zhou, Zongwei and Rahman Siddiquee, Md Mahfuzur and Tajbakhsh, Nima and Liang, Jianming},
  booktitle={International workshop on deep learning in medical image analysis},
  pages={3--11},
  year={2018},
  organization={Springer}
}

@article{liu2024wsi,
  title={WSI-SAM: Multi-resolution Segment Anything Model (SAM) for histopathology whole-slide images},
  author={Liu, Hong and Yang, Haosen and van Diest, Paul J and Pluim, Josien PW and Veta, Mitko},
  journal={arXiv preprint arXiv:2403.09257},
  year={2024}
}

@misc{rostami2025segmentcrackdeepsemantic,
      title={Segment Any Crack: Deep Semantic Segmentation Adaptation for Crack Detection}, 
      author={Ghodsiyeh Rostami and Po-Han Chen and Mahdi S. Hosseini},
      year={2025},
      eprint={2504.14138},
      archivePrefix={arXiv},
      primaryClass={cs.CV},
      url={https://arxiv.org/abs/2504.14138}, 
}

@inproceedings{zhaotuning,
  title={Tuning LayerNorm in Attention: Towards Efficient Multi-Modal LLM Finetuning},
  year = {2023},
  author={Zhao, Bingchen and Tu, Haoqin and Wei, Chen and Mei, Jieru and Xie, Cihang},
  booktitle={The Twelfth International Conference on Learning Representations}
}

@inproceedings{zhang20252dmamba,
  title={2dmamba: Efficient state space model for image representation with applications on giga-pixel whole slide image classification},
  author={Zhang, Jingwei and Nguyen, Anh Tien and Han, Xi and Trinh, Vincent Quoc-Huy and Qin, Hong and Samaras, Dimitris and Hosseini, Mahdi S},
  booktitle={Proceedings of the Computer Vision and Pattern Recognition Conference},
  pages={3583--3592},
  year={2025}
}

@article{gadermayr2024multiple,
  title={Multiple instance learning for digital pathology: A review of the state-of-the-art, limitations \& future potential},
  author={Gadermayr, Michael and Tschuchnig, Maximilian},
  journal={Computerized Medical Imaging and Graphics},
  volume={112},
  pages={102337},
  year={2024},
  publisher={Elsevier}
}

@article{dippel2024rudolfv,
  title={RudolfV: a foundation model by pathologists for pathologists},
  author={Dippel, Jonas and Feulner, Barbara and Winterhoff, Tobias and Milbich, Timo and Tietz, Stephan and Schallenberg, Simon and Dernbach, Gabriel and Kunft, Andreas and Heinke, Simon and Eich, Marie-Lisa and Julika Ribbat-Idel and Rosemarie Krupar and Philipp Anders and Niklas Prenißl and Philipp Jurmeister and David Horst and Lukas Ruff and Klaus-Robert Müller and Frederick Klauschen and Maximilian Alber},
  journal={arXiv preprint arXiv:2401.04079},
  year={2024}
}

@misc{labelbox,
  author       = {Labelbox, Inc.},
  title        = {Labelbox: data labeling platform for AI},
  howpublished = {\url{https://labelbox.com}},
  year         = {2025},
  note         = {Accessed 10 May 2025}
}

@article{EntropyMasker,
  title={An automatic entropy method to efficiently mask histology whole-slide images},
  author={Song, Yipei and Cisternino, Francesco and Mekke, Joost M and de Borst, Gert J and de Kleijn, Dominique PV and Pasterkamp, Gerard and Vink, Aryan and Glastonbury, Craig A and van der Laan, Sander W and Miller, Clint L},
  journal={Scientific Reports},
  volume={13},
  number={1},
  pages={4321},
  year={2023},
  publisher={Nature Publishing Group UK London}
}

@article{ravi2024sam,
  title={Sam 2: Segment anything in images and videos},
  author={Ravi, Nikhila and Gabeur, Valentin and Hu, Yuan-Ting and Hu, Ronghang and Ryali, Chaitanya and Ma, Tengyu and Khedr, Haitham and R{\"a}dle, Roman and Rolland, Chloe and Gustafson, Laura and  Eric Mintun and Junting Pan and Kalyan Vasudev Alwala and Nicolas Carion and Chao-Yuan Wu and Ross Girshick and Piotr Dollár and Christoph Feichtenhofer},
  journal={arXiv preprint arXiv:2408.00714},
  year={2024}
}

@article{jaume2024hest,
  title={Hest-1k: A dataset for spatial transcriptomics and histology image analysis},
  author={Jaume, Guillaume and Doucet, Paul and Song, Andrew and Lu, Ming Yang and Almagro P{\'e}rez, Cristina and Wagner, Sophia and Vaidya, Anurag and Chen, Richard and Williamson, Drew and Kim, Ahrong and Faisal Mahmood},
  journal={Advances in Neural Information Processing Systems},
  volume={37},
  pages={53798--53833},
  year={2024}
}

@article{meirelles2022effective,
  title={Effective active learning in digital pathology: A case study in tumor infiltrating lymphocytes},
  author={Meirelles, Andr{\'e} LS and Kurc, Tahsin and Saltz, Joel and Teodoro, George},
  journal={Computer Methods and Programs in Biomedicine},
  volume={220},
  pages={106828},
  year={2022},
  publisher={Elsevier}
}

@article{xiong2025survey,
  title={A survey of pathology foundation model: Progress and future directions},
  author={Xiong, Conghao and Chen, Hao and Sung, Joseph JY},
  journal={arXiv preprint arXiv:2504.04045},
  year={2025}
}

@article{huang2023visual,
  title={A visual--language foundation model for pathology image analysis using medical twitter},
  author={Huang, Zhi and Bianchi, Federico and Yuksekgonul, Mert and Montine, Thomas J and Zou, James},
  journal={Nature medicine},
  volume={29},
  number={9},
  pages={2307--2316},
  year={2023},
  publisher={Nature Publishing Group US New York}
}

@article{lu2024visual,
  title={A visual-language foundation model for computational pathology},
  author={Lu, Ming Y and Chen, Bowen and Williamson, Drew FK and Chen, Richard J and Liang, Ivy and Ding, Tong and Jaume, Guillaume and Odintsov, Igor and Le, Long Phi and Gerber, Georg and Anil V. Parwani and Andrew Zhang and Faisal Mahmood},
  journal={Nature medicine},
  volume={30},
  number={3},
  pages={863--874},
  year={2024},
  publisher={Nature Publishing Group US New York}
}

@article{chen2024towards,
  title={Towards a general-purpose foundation model for computational pathology},
  author={Chen, Richard J and Ding, Tong and Lu, Ming Y and Williamson, Drew FK and Jaume, Guillaume and Song, Andrew H and Chen, Bowen and Zhang, Andrew and Shao, Daniel and Shaban, Muhammad and Mane Williams and Lukas Oldenburg and Luca L. Weishaupt and Judy J. Wang and Anurag Vaidya and Long Phi Le and Georg Gerber and Sharifa Sahai and Walt Williams and Faisal Mahmood},
  journal={Nature medicine},
  volume={30},
  number={3},
  pages={850--862},
  year={2024},
  publisher={Nature Publishing Group US New York}
}

@inproceedings{abmil,
  title={Attention-based deep multiple instance learning},
  author={Ilse, Maximilian and Tomczak, Jakub and Welling, Max},
  booktitle={International conference on machine learning},
  pages={2127--2136},
  year={2018},
  organization={PMLR}
}

@inproceedings{dsmil,
  title={Dual-stream multiple instance learning network for whole slide image classification with self-supervised contrastive learning},
  author={Li, Bin and Li, Yin and Eliceiri, Kevin W},
  booktitle={Proceedings of the IEEE/CVF conference on computer vision and pattern recognition},
  pages={14318--14328},
  year={2021}
}

@inproceedings{dtfd,
  title={Dtfd-mil: Double-tier feature distillation multiple instance learning for histopathology whole slide image classification},
  author={Zhang, Hongrun and Meng, Yanda and Zhao, Yitian and Qiao, Yihong and Yang, Xiaoyun and Coupland, Sarah E and Zheng, Yalin},
  booktitle={Proceedings of the IEEE/CVF conference on computer vision and pattern recognition},
  pages={18802--18812},
  year={2022}
}

@inproceedings{rrt,
  title={Feature re-embedding: Towards foundation model-level performance in computational pathology},
  author={Tang, Wenhao and Zhou, Fengtao and Huang, Sheng and Zhu, Xiang and Zhang, Yi and Liu, Bo},
  booktitle={Proceedings of the IEEE/CVF conference on computer vision and pattern recognition},
  pages={11343--11352},
  year={2024}
}

@article{transmil,
  title={Transmil: Transformer based correlated multiple instance learning for whole slide image classification},
  author={Shao, Zhuchen and Bian, Hao and Chen, Yang and Wang, Yifeng and Zhang, Jian and Ji, Xiangyang and Yongbing Zhang},
  journal={Advances in neural information processing systems},
  volume={34},
  pages={2136--2147},
  year={2021}
}

@inproceedings{wikg,
  title={Dynamic graph representation with knowledge-aware attention for histopathology whole slide image analysis},
  author={Li, Jiawen and Chen, Yuxuan and Chu, Hongbo and Sun, Qiehe and Guan, Tian and Han, Anjia and He, Yonghong},
  booktitle={Proceedings of the IEEE/CVF conference on computer vision and pattern recognition},
  pages={11323--11332},
  year={2024}
}

@inproceedings{shaomultiple,
  title={Do Multiple Instance Learning Models Transfer?},
  author={Shao, Daniel and Chen, Richard J and Song, Andrew H and Runevic, Joel and Lu, Ming Y and Ding, Tong and Mahmood, Faisal},
  booktitle={Forty-second International Conference on Machine Learning}
}

@article{hosseini2026efficient,
  title={Efficient self-supervised Barlow twins from limited tissue slide cohorts for colonic pathology diagnostics},
  author={Hosseini, Mahdi S and Notton, Cassandre and Sharma, Vasudev and Trinh, Vincent Quoc-Huy and Chen, Lina and Varma, Sonal and Xu, Minqi},
  journal={Medical Image Analysis},
  pages={104004},
  year={2026},
  publisher={Elsevier}
}

@article{li2025diagnostic,
  title={Diagnostic text-guided representation learning in hierarchical classification for pathological whole slide image},
  author={Li, Jiawen and Sun, Qiehe and Yan, Renao and Wang, Yizhi and Fu, Yuqiu and Wei, Yani and Guan, Tian and Shi, Huijuan and He, Yonghong and Han, Anjia},
  journal={Medical Image Analysis},
  pages={103894},
  year={2025},
  publisher={Elsevier}
}

\end{document}